\documentclass[fleqn,usenatbib]{mnras}

\usepackage{newtxtext,newtxmath}

\usepackage[T1]{fontenc}
\usepackage{ae,aecompl}
\usepackage[nolist]{acronym}
\usepackage{trackchanges}

\usepackage{graphicx}	
\usepackage{amsmath}	
\usepackage{makecell}
\usepackage{xcolor}
\usepackage{comment}
\usepackage{acronym}
\usepackage{xurl}

\newcommand{\cxo}{{\it Chandra}}

\newcommand{\ergcm}[1]{erg~cm$^{-2}$ s$^{-1}$}
\newcommand{\ngc}{NGC~2663}
\newcommand{\Swift}{{\it Swift}}

\newcommand{\D}{$^\circ$}

\newcommand{\FRI}{FR\,{\sc i}}
\newcommand{\FRII}{FR\,{\sc ii}}

\begin{acronym}

\acro{AGN}{Active Galactic Nuclei}
\acro{ATCA}{Australia Telescope Compact Array}
\acro{ASKAP}{Australian Square Kilometre Array Pathfinder}
\acro{EMU}{Evolutionary Map of the Universe}
\acro{FWHM}{Full Width at Half-Maximum power}
\acro{HST}{Hubble Space Telescope}
\acro{ISM}{Interstellar Medium}
\acro{HIPASS}{H\,{\sc i} Parkes All Sky Survey}
\acro{MIRIAD}[{\tt MIRIAD}]{Multichannel Image Reconstruction, Image Analysis and Display}
\acro{MWA}{Murchison Widefield Array}
\acro{NRAO}{National Radio Astronomy Observatory}
\acro{NVSS}{NRAO VLA Sky Survey}
\acro{pc}{parsec\acroextra{: 1 pc $\simeq 3.09 \times 10^{16}$ m}}
\acro{RFI}{Radio-Frequency Interference}
\acro{SI}[$\alpha$]{Spectral Index\acroextra{, $S \propto \nu^\alpha$}}
\acro{SMBH}{Super Massive Black Hole}
\acro{BH}{Black Hole}
\acrodefplural{BH}{Black Holes}
\acro{WISE}{Wide-Field Infrared Survey Explorer}
\acro{VLBI}{Very Long Baseline Interferometry} 
\acro{VLBA}{Very Long Baseline Array}

\end{acronym}

\title{Collimation of the kiloparsec-scale radio jets in \ngc}

\author[V. Velovi\'c et al.]{Velibor Velovi\'{c},$^{1}$\thanks{E-mail: v.velovic@westernsydney.edu.au}
M. D. Filipovi\'{c},$^{1}$
L. Barnes,$^{1}$
R. P. Norris,$^{1,2}$
C. D. Tremblay,$^{3,4}$
\newauthor
G. Heald,$^{4}$
L. Rudnick,$^{5}$ 
S. S. Shabala,$^{6}$
T. G. Pannuti,$^{7}$
H. Andernach,$^{8}$
O. Titov,$^{9}$
\newauthor
S. G. H. Waddell,$^{10}$
B. S. Koribalski,$^{2,1}$
D. Grupe,$^{11}$
T. Jarrett,$^{12}$
R. Z. E. Alsaberi,$^{1}$
\newauthor
E. Carretti,$^{13}$
J. D. Collier,$^{14,1,4}$
S. Einecke,$^{14}$
T. J. Galvin,$^{16, 4}$
A. Hotan,$^{4}$
P. Manojlovi\'{c},$^{1}$
\newauthor
J. Marvil,$^{17}$
K. Nandra,$^{10}$
T. H. Reiprich,$^{18}$
G. Rowell,$^{14}$
M. Salvato,$^{10}$
M. Whiting $^{2}$\\
$^{1}$School of Science, Western Sydney University, Locked Bag 1797, Penrith South DC, NSW 2751, Australia\\
$^{2}$CSIRO Space and Astronomy, Australia Telescope National Facility, PO Box 76, Epping NSW 1710, Australia\\
$^{3}$SETI Institute, Mountain View, CA 94043, USA\\
$^{4}$CSIRO Space and Astronomy, Australia Telescope National Facility, PO Box 1130, Bentley WA 6102, Australia\\
$^{5}$School of Physics and Astronomy, University of Minnesota, Minneapolis, MN 55455, USA \\ 
$^{6}$School of Natural Sciences, Private Bag 37, University of Tasmania, Hobart, TAS 7001, Australia \\
$^{7}$Physics, Earth Science, and Space System Engineering, Morehead State University, Martindale Drive, Morehead, KY 40351, USA\\ 
$^{8}$Departmento de Astronomía, DCNE, Universidad de Guanajuato, Callej\'on de Jalisco s/n, Guanajuato, C.P. 36023, GTO, Mexico\\
$^{9}$Geoscience Australia, Canberra, ACT 2601, Australia \\
$^{10}$Max-Planck-Institut für extraterrestrische Physik, Gießenbachstraße, D-85748 Garching, Germany\\
$^{11}$Department of Physics, Geology, and Engineering Technology, Northern Kentucky University, 1 Nunn Dr. Highland Heights, KY 41099, USA\\
$^{12}$Department of Astronomy, University of Cape Town, Rondebosch, South Africa \\
$^{13}$INAF -- Istituto di Radioastronomia, Via P. Gobetti 101, 40129, Bologna, Italy\\
$^{14}$School of Physical Sciences, The University of Adelaide, Adelaide 5005, Australia \\
$^{15}$The Inter-University Institute for Data Intensive Astronomy (IDIA), Department of Astronomy, University of Cape Town, Private Bag X3, Rondebosch, 7701, South Africa\\
$^{16}$International Centre for Radio Astronomy Research, Curtin University, Bentley, WA 6102, Australia\\
$^{17}$National Radio Astronomy Observatory, PO Box 0, Socorro, NM87801, USA\\
$^{18}$Argelander-Institut für Astronomie (AIfA), Universität Bonn, Auf dem H\"ugel 71, 53121 Bonn, Germany \\
}

\date{Accepted XXX. Received YYY; in original form ZZZ}

\pubyear{2022}

\begin{document}
\label{firstpage}
\pagerange{\pageref{firstpage}--\pageref{lastpage}}
\maketitle

\begin{abstract}

We present the discovery of highly-collimated radio jets spanning a total of 355~kpc around the nearby elliptical galaxy \ngc, and the possible first detection of recollimation on kiloparsec scales. The small distance to the galaxy ($\sim$28.5~Mpc) allows us to resolve portions of the jets to examine their structure. We combine multiwavelength data: radio observations by the \ac{MWA}, the \ac{ASKAP} and the \ac{ATCA}, and X-ray data from \cxo, \Swift\ and SRG/eROSITA. We present intensity, rotation measure, polarisation, spectral index and X-ray environment maps. Regions of the southern jet show simultaneous narrowing and brightening, which can be interpreted as a signature of the recollimation of the jet by external, environmental pressure, though it is also consistent with an intermittent \ac{AGN} or complex internal jet structure. X-ray data suggest that the environment is extremely poor; if the jet is indeed recollimating, the large recollimation scale (40~kpc) is consistent with a slow jet in a low-density environment. 

\end{abstract}


\section{Introduction} 
\ac{AGN} jets are highly-collimated outflows of
relativistic plasma, generated from the accretion of material onto a \ac{SMBH} in a galactic centre. SMBHs are believed to exist in all elliptical galaxies and spiral galaxy bulges \citep{1995ARA&A..33..581K,1998AJ....115.2285M}. Relativistic jets launched from SMBHs have been observed on scales from parsecs to megaparsecs, interacting with cosmic environments from the immediate neighbourhood of the accretion disk to the intracluster and intergalactic medium.

Understanding \ac{AGN} jets is crucial for characterising the effect of \acp{BH} on the formation of structure in the Universe, especially the high-mass end of the stellar-mass function. For galaxies in dark matter haloes more massive than $10^{12}$ M$_\odot$, heat and outflowing material from \ac{AGN} is believed to be the most important form of feedback against baryon collapse and star formation \citep{2019MNRAS.490.3234N,2006MNRAS.365...11C}.

How do \ac{AGN} jets interact with their cosmic environment? A range of physical processes affect the jet's dynamics and stability. As the initially over-pressured jet streams away from the centre of the galaxy, its pressure decreases with distance from the jet base. Eventually it reaches the critical point where the jet pressure falls below the pressure of the external environment. The pressure mismatch can drive an oscillation in the width of the jet, causing it to narrow and brighten with a characteristic length scale. This phenomenon is called \emph{recollimation} (see, e.g. \citealp{2015ApJ...809...38M,1997ApJ...482L..33G,1997MNRAS.288..833K}). A parsec-scale jet recollimation is suggested in both, distant ($z=6.1$) blazar PSO~J030947.49+271757.31 \citep{2020A&A...643L..12S} as well as in nearby galaxies M\,87 \citep{2014ApJ...781L...2A} and 1H~0323+342 \citep{2018ApJ...860..141H}. 

The study of recollimation phenomena, then, can shed light on both the origin of jets, and the impact of \ac{AGN} on galaxy formation. There are (at least) two scenarios that could produce recollimation. The first involves an under-pressured jet, relative to the surrounding medium. The second scenario involves a steep drop in the ambient pressure \citep{1988ApJ...334..539D}. 

\begin{figure*}
\centering
\includegraphics[width=\textwidth,angle=0]{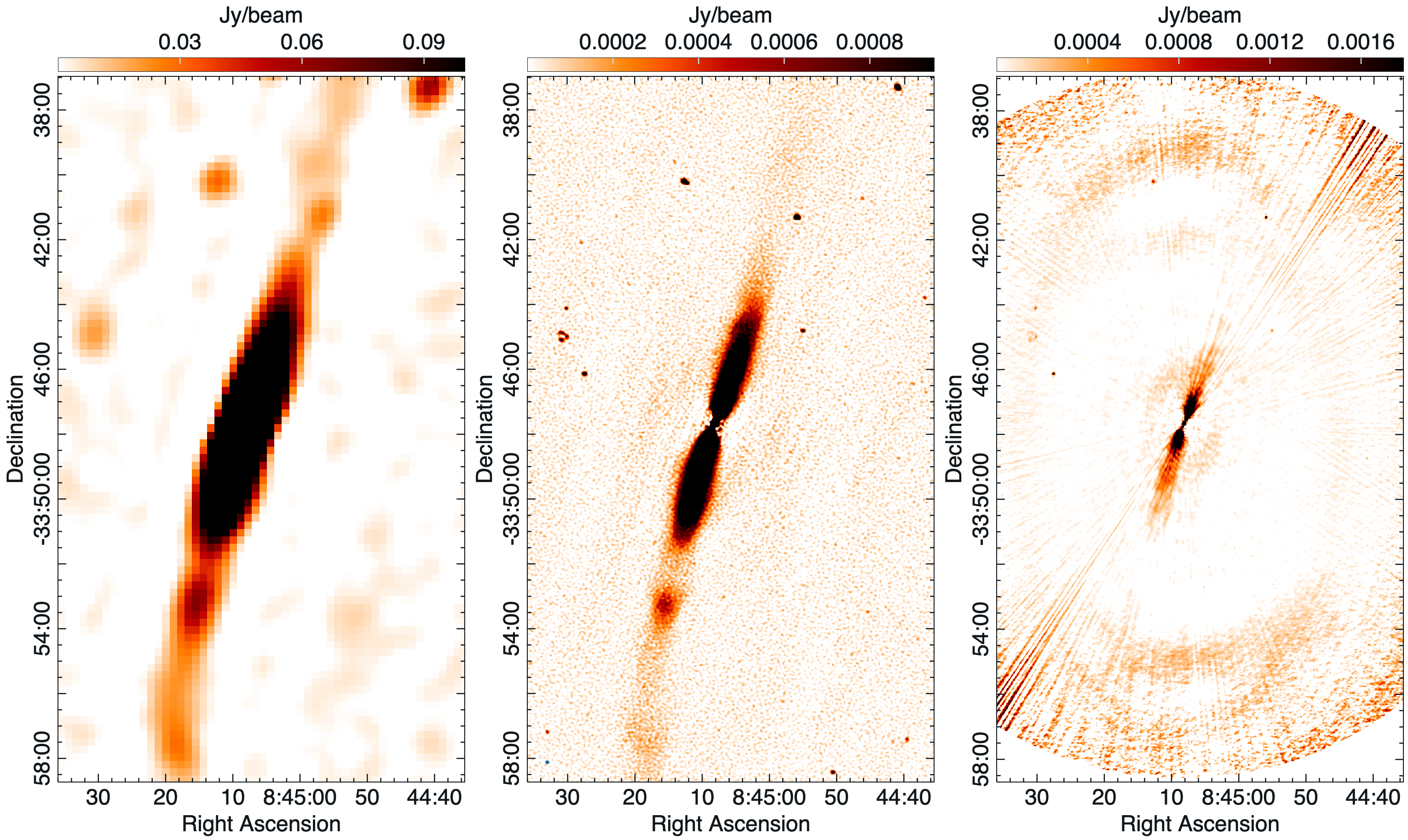}
\caption{ \ngc\ images of \ac{MWA} centered at 200 MHz shown on the left and \ac{ASKAP} at 1520\,MHz in the \emph{Middle} show potential recollimation knots in the southern jet. \emph{Right:} High frequency ATCA image centered at 5500\,MHz did not detect anything in that region.
}
\label{figure:multi_Freq_ngc2663}
\end{figure*}


\ngc\ is an elliptical galaxy at a distance of 28.5~Mpc \citep[$z=0.0067$,][]{1997ApJS..109..333W}, whose jets span 355~kpc from one end to the other.  This makes it one of the largest radio galaxies (in projected angular size) in the nearby Universe ($\sim200$~Mpc), with absolute optical magnitude of $-19.94$. \ngc\ contains a gaseous disc, but does not have an obvious stellar disc. \ngc\ has a compact central radio source (see, e.g. \citealp{1983MNRAS.202..703D,1989MNRAS.240..591S}). The velocity of gas rotation is very low, and stellar rotation is not detected \citep{2014MNRAS.440.2419R}. Using the optical spectral classification from \citet{1987ApJS...63..295V}, we find that optical spectral emission line ratios obtained by \cite{2014MNRAS.440.2442R} strongly suggest that \ngc\ contains an \ac{AGN}. It exhibits a large metallicity gradient, and the derived (Mg/Fe) abundance ratio is higher than Solar. Na~D lines have not been detected, which indicates that the light from its nucleus is not significantly affected by the \ac{ISM} \citep{2014MNRAS.440.2419R}. 

Here, we present a multi-wavelength study of \ngc\ from radio to X-ray, seeking to understand the jet and its effect on its galactic environment. Section~\ref{Sect:data} describes the instruments and data. In Section~\ref{Sect:results} we present our observational results, divided into six subsections: jet structure, polarisation, spectral index, environment, host galaxy, and X-ray emission. In Section~\ref{sect:DC} we discuss theoretical models that can explain our observations of \ngc. Section \ref{distance}, in anticipation of future recollimation candidates in the \ac{EMU} survey, discusses how \ngc\ would appear if it were observed at larger distances.

\section{Observations and Data}
 \label{Sect:data}
 
In this section, we present observations of \ngc\ at radio wavelengths from \ac{MWA} at 200~MHz, \ac{ASKAP} at 1520~MHz, \ac{ATCA} at 1384~MHz and 2368~MHz, the \ac{HIPASS}, and the \ac{VLBA} at 2300~MHz and 8400~MHz. We also present data at infrared wavelengths from the \ac{WISE}, at optical wavelengths from the \ac{HST}, at UV wavelengths from \Swift, and in the X-ray regime from \cxo, eROSITA and \Swift. A summary of the observations is given in Table~\ref{Obs_Summary}.

\subsection{Murchison Widefield Array}
\label{Sect:MWA2}

Observations were carried out using the \ac{MWA} \citep{2013PASA...30....7T}, located at the Murchison Radio-astronomy Observatory, as part of the GaLactic and Extragalactic All-sky MWA eXtended (GLEAM-X) survey (Hurley-Walker et al. in prep). At the time of these observations, the telescope had 128 dipole tiles (110 online during observations), spread across 5.5~km (Phase~II; \citealt{Wayth-PhaseII}), with a primary beam of 30~degrees \ac{FWHM} and a synthesized beam of $\approx$1~arcmin \ac{FWHM} at 200\,MHz.

The \ac{MWA} uses a two-stage polyphase filter bank to channelize the data. The first stage separates the 30.72~MHz bandwidth into 24$\times$1.28~MHz coarse channels while the second stage breaks up each coarse channel into 128$\times$10~kHz fine channels. GLEAM-X observed in a series of drift scans, one per night, iterating through each 30.72-MHz channel between 72--231\,MHz, changing every two minutes. For this work, we used observations taken over UTC2018-02-04~15:30--16:38, UTC2018-02-20~14:27--15:35, and UTC2018-05-03~10:04--10:52  at 170--200 and 200--231\,MHz, comprising 112\,minutes of integration, but at heterogeneous sensitivity with the source appearing in different regions of the primary beam.

The data were calibrated on a sky model based on GLEAM ExGal \citep{2017MNRAS.464.1146H}, and further refined by using one round of self-calibration on an initial \textsc{clean} using \textsc{WSClean} \citep{2014MNRAS.444..606O}. The final imaging was performed with a Briggs' ''robust'' weighting of 0.0 \citep{1995AAS...18711202B}, forming a mask at $3\times$ the local root-mean-square noise of the residuals $\sigma$, then cleaning down to $1\sigma$ inside the mask, and producing a multi-frequency synthesis image across each 30.72\,MHz band. The 200--231\,MHz images were convolved to the same angular resolution as the 170--200\,MHz images ($0.8'\times1'$) and combined using the mosaicking software \textsc{swarp} \citep{2002ASPC..281..228B}. The final flux density scale was tested against that of GLEAM ExGal for sources with signal-to-noise$>30$, and the median integrated flux density ratio was found to be $0.91\pm0.06$.

\subsection{The Australian Square Kilometre Array Pathfinder}
\label{Sect:ASKAP}

The \ac{ASKAP} \citep{2008ExA....22..151J,ASKAP_System} is located at the Murchison Radio-astronomy Observatory in Western Australia. Each of the 36 antennas is equipped with a Phased Array Feed (PAF) \citep{Schinckel-PAF} giving it a field of view of $\approx 25$~deg$^2$. Our data were taken with the Square {\sc 6$\times$6} footprint, a pitch of 0.9, and a footprint rotation of 45~degrees. They were observed on 10$^\mathrm{th}$~April 2019 at a central frequency of 1520.5~MHz with 288~MHz of bandwidth in continuum mode (1~MHz channel widths) using a mosaic of four dithered exposures around \ngc\ with resolution of $ 6.0 \times 5.1 $~arcsec. All data were processed using \ac{ASKAP}{\sc soft} \citep{Guzman_Askapsoft} for calibration and imaging.

The observations were taken as part of the commissioning of the high-band of \ac{ASKAP} \citep{McConnell-ASKAP} and 23 antennas were online with a maximum baseline of 6\,km. They were taken in full polarisation, and observations of PKS\,B1934--638 were performed immediately adjacent in time to the target field for purposes of instrumental calibration. The calibration observation contained one calibrator scan of duration five minutes at the centre of each of the 36~PAF beams.

We performed a first order image-plane correction for the leakage in all Stokes parameters. After that correction on top of mosaicking on the basis of the brightest unpolarised source near \ngc, it appears that the Stokes I leakage can be constrained to $<$0.7$\pm$0.2~per~cent. This is consistent with results from other \ac{ASKAP} imaging work done within the POSSUM project (i.e. \citealt{2021arXiv210201702A}). We rejected the mid-band channels affected by \ac{RFI} and then created a rotation measure (RM) synthesis map \citep{2005A&A...441.1217B}. 

\subsection{Australia Telescope Compact Array}
\label{Sect:ATCA}

We used \ac{ATCA} observational data from project C3370 that were taken on the 30$^\mathrm{th}$ September 2020 in the 6B array configuration, with the shortest baseline between two antennas of $214$~m. These observations used a bandwidth of 2048~MHz centred at frequencies of 5500~MHz and 9000~MHz, for which the largest well-imaged structure have the angular sizes of $80$~arcsec and $44$~arcsec respectively.

The observations totalled $\sim$12~hours of integration time in each band. PKS\,B1934--638 was used as the primary (absolute flux density and bandpass) calibrator, and PKS\,B0826--373 was used as the secondary (time-varying gains) calibrator. The \textsc{miriad}\footnote{http://www.atnf.csiro.au/computing/software/miriad/} \citep{1995ASPC...77..433S} and \textsc{karma}\footnote{http://www.atnf.csiro.au/computing/software/karma/} \citep{1995ASPC...77..144G} software packages were used to reduce and analyze the data. 
Images were formed using \textsc{miriad} and the multi-frequency synthesis tasks \citep{1994A&AS..108..585S} with a Briggs weighting of robust (R=0) parameter for both the 5500 and 9000~MHz images. Both images were deconvolved using \textsc{mfclean} and primary beam correction was applied afterwards. 
The 5000~MHz image has a resolution of $4\times4$~arcsec, while the 9000~MHz image has a resolution of $2\times2$~arcsec. We also note an effect of the primary beam correction and the $uv$ limitation in the outer region the \ac{ATCA} images.

\begin{table*}
\begin{center}
\caption{Telescopes and observation details used for \ngc.}
\begin{tabular}{ l l c c c c c } 
\hline
    Radio Observatory & Date &\makecell{Frequency \\(MHz)}  & \makecell{Bandwidth \\ (MHz)} & \makecell{FoV \\(degrees)} & \makecell{Resolution \\ (arcsec)} &\makecell{ P.A.  \\ (degrees)} \\ 
\hline
    \ac{ASKAP} & 2019~Apr~10 & $1520$ & $288$ & $25$ &$ 6.0 \times 5.1 $ & $79.4$ \\ 
    ATCA &2020~Sep~30& $5500 $ & $ 2048 $& $0.4$ & $15.6 \times 6.7$ & $-6.5 $ \\
    ATCA &2020~Sep~30& $9000 $ & $2048 $& $0.4$ & $9.1\times3.6$  & $ -5.0$   \\
    \ac{MWA}  & 2018~Feb~4, 20;~Mar~5  & $ 200$ & $61.44$ & $30$ & $64.4\times49.0$ & $151.6$\\
    \ac{VLBA}  & $2009-2018$ & $2300$ \& $8400$& $32$ & $0.005$ & $<0.004$ & $-28.3$\\
\hline
    X-ray  observatory & Date & \makecell{ Observing time \\ (Sec) } & \makecell{Energy band \\ (keV)} & \makecell{FoV \\(arcmin)} & \makecell{Resolution \\ (arcsec)}   \\
\hline
    {\it Chandra}  & 2011~Jul~12 & $7500$ & $0.5-7$ & $30 \times 30$ & $0.5$   \\
    eROSITA / & 2020~May~11-18 & $262$ & $0.1-8$  & $25$  & $15$  \\
    \Swift\  & 2018~Sep~2,16 & $750$, $483$ &$ 0.3-10$ & $23.6 \times 23.6$ & $18$  \\
\hline
\hline
\label{Obs_Summary}
\end{tabular}
\end{center}
\end{table*}
\subsection{Very Long Baseline Array}

The galaxy \ngc\ has been observed by the \ac{VLBA} operated by the National Radio Astronomical Observatory (NRAO). The \ac{VLBA} is a network of ten 25-meter radio telescopes located across the United States \citep{Napier1994}. This network regularly participates in the observing program organised by the International \ac{VLBI} Service (IVS) \citep{Schuh2012}. The goal of this project is to measure high-precision absolute positions of weak radio sources. Observations are done in dual-frequency mode (S-band, 2300~MHz, and X-band, 8400~MHz), but S-band is used only to calibrate very fast fluctuations of the ionosphere. There are five IVS experiments between 2009 and 2018 in which the galaxy \ngc\ was successfully detected as IERS B0843--336  (Table \ref{table_VLBI_obs}) -- RDV75 (2009~May~13), UF001H (1$^{\rm st}$~May~2017), UF001S (9$^{\rm th}$~October~2017), UF001T (21$^{\rm st}$~October~2017), UG002F (29$^{\rm th}$~April~2018). Where RDV stands for ``Research and Development \ac{VLBI}'', while the other letters (UF and UG) are internal designations for \ac{VLBA} experiments. Numbers 001, 002 stand for proposal term, and the following  letters (A, B, ..., H) are the serial number of an experiment.
Correlated \ac{VLBI} observations available at the IVS data were processed by the {\sc OCCAM} software \citep{Titov2004} using the standard procedure of the \ac{VLBI} data reduction recommended by the International Earth Rotation Service (IERS) \citep{iers2010}. 

The positions of 303 so-called ICRF3 defining radio sources were fixed \citep{2020A&A...644A.159C}, and positions of all other sources were estimated in each of these five experiments separately. The coordinate estimates for the peak brightness position of \ngc\ vary within 4~mas over these five epochs, while the formal $1\sigma$ error is between 0.7 and 1.7~mas in RA, and between 2 and 5~mas in Dec.
We thus conclude that the observed positional variations are below the $3\sigma$ threshold, so the radio source \ngc\ is astrometrically stable. 

\subsection{Hubble Space Telescope}

We have also accessed \ac{HST} observations from the Hubble Legacy Archive (Proposal ID 11219). These observations are from 30$^{\rm th}$~October~2007 and the total exposure time is 1151.84~seconds. The wide-field camera~3 $H$ is used with the IR Wide-Band (W) Filter F160W that covers 1400--1700~nm range and effective resolution is 0.3~arcsec.

\begin{table}
\caption{\ac{VLBA} observations of \ngc. }
\begin{tabular}{l l l l l}
 \hline
 Date & \makecell{RA (J2000) \\(h m s)} & \makecell{$\sigma_{\rm RA}$ \\ (sec)} & \makecell{Dec. (J2000) \\  (\D\ \arcmin\ \arcsec)} & \makecell{$\sigma_{\rm Dec}$ \\ (\arcsec)} \\ [1ex] 
 \hline \hline
      2009~May~13 & $08\, 45\, 08.144446$ & $0.000082$ & $-33\, 47\, 41.0636$ & $0.0046$ \\ [1ex] 
      2017~May~1  & $08\, 45\, 08.144501$ & $0.000047$ & $-33\, 47\, 41.0666$ & $0.0020$ \\ [1ex] 
      2017~Oct~9  & $08\, 45\, 08.144585$ & $0.000111$ & $-33\, 47\, 41.0678$ & $0.0035$ \\ [1ex] 
      2017~Oct~21 & $08\, 45\, 08.144428$ & $0.000081$ & $-33\, 47\, 41.0639 $& $0.0031$ \\ [1ex]
      2018~Apr~29 & $08\, 45\, 08.144510$ & $0.000056$ & $-33\, 47\, 41.0655$ & $0.0032$ \\ [1ex]
 \hline
\label{table_VLBI_obs}
\end{tabular}
\end{table}

\subsection{Chandra X-Ray Observatory}

\ngc\ was the target of a pointed observation made with the \cxo\ X-ray Observatory on 12$^{\rm th}$~July~2011. The observation was made using the ACIS-S3 chip in very faint mode. The archival dataset was downloaded from the \cxo\ data archive\footnote{http://cda.harvard.edu} and processed using the CIAO Interactive Analysis of Observations (CIAO) software package \citep{2006SPIE.6270E..1VF} Version 4.12 (CALDB Version 4.9.1). The CIAO tool \textsc{chandra\_repro} was used to reprocess the dataset with the latest calibration: in addition, standard light curve filtering to address the effects of any background  flares was applied to the dataset. After this processing, the effective exposure time of the observation was $\sim$ 7.5~kiloseconds. The CIAO tool \textsc{dmxtract} was used to extract a spectrum and a light curve of the central nuclear source while the CIAO tool \textsc{fluximage} was used to produce exposure-corrected images of the central source. The results of our imaging and spectroscopic analysis of this observation are presented in Subsection \ref{chandra}.

\subsection{Neil Gehrels Swift Observatory}
\label{swift_obs}

The Neil Gehrels Gamma-Ray Burst Explorer Mission \Swift\ \citep{2004ApJ...611.1005G} observed \ngc\ on 2$^{\rm nd}$ and 16$^{\rm th}$ September~2018 for 750 and 483~sec, respectively (target ID 3109192). The observations with the \Swift\ X-ray Telescope \citep[XRT,]{2005SSRv..120..165B} were obtained in photon counting mode \citep[pc mode,][]{2004SPIE.5165..217H}.  Source counts were extracted with {\it xselect} in a circular region with a radius of 47~arcsec, and background counts in a nearby source-free region with a radius of 236~arcsec. Due to the small number of counts (16 in total) we combined the data of two observations into one source and background spectrum. The auxiliary response files were created from each source spectrum with {\it xrtmkarf} and combined into one with the ftool {\it addarf}. For the spectral analysis we use Cash statistics \citep{1979ApJ...228..939C} within XSPEC version 12.10.1f \citep{1996ASPC..101...17A}.

\Swift\ also observed the field of \ngc\ with its UV-Optical Telescope \citep[UVOT,][]{ 2005SSRv..120...95R} in its u and uvw1 filters during each observation (central wavelengths are 3465\AA\ and 2600\AA, respectively). The magnitudes and fluxes were determined with the UVOT tool {\it uvotsource} with a source extraction radius of 3~arcsec and the aperture was corrected by the aperture parameter set to  {\it apercorr=curveofgrowth}. The most recent calibration files were used as described in \cite{2008MNRAS.383..627P} and \cite{2010MNRAS.406.1687B}. The magnitudes were corrected for Galactic reddening ($E_{\rm B-V}=0.312$) following equation (2) in \cite{2009ApJ...690..163R} using the reddening curves by \cite{1989ApJ...345..245C}. The results of \Swift\ observations are presented in Subsection \ref{swift_res}.

\subsection{eROSITA}
\label{eRosita_data_proc}

\ngc\ was observed with eROSITA (\citealt{Merloni+2012,Predehl+2020}), the primary instrument onboard the Russian \textit{Spektrum-R\"{o}ntgen-Gamma} (SRG) mission as part of the first all-sky survey (eRASS1). The observation took place between 11$^{\rm th}$ and 18$^{\rm th}$~May~2020, and all seven cameras (TM1-TM7) were operational. Version \textit{c946} of the standard eROSITA data processing \citep{2021arXiv210614517B} was used to retrieve prepared event lists. Source photons were extracted from a 60~arcsec region centred on the source, while background photons were extracted from an annulus with an inner radius of 140~arcsec and an outer radius of 240~arcsec. There are 48 counts from the source detected in the $0.1-8~{\rm keV}$ energy range. The source and background spectra were grouped to a minimum of one count per bin, and Cash statistics \citep{1979ApJ...228..939C} are used in spectral analysis. The background has been modelled, rather than subtracted from the source. 
Complete analysis and results from eROSITA are presented in Subsection \ref{erosita}.

\section{Results}
\label{Sect:results}

\ngc\ is a large and massive elliptical galaxy with stellar mass of $5.8~\times~10^{11} \textrm{M}_\odot$ \citep{2013AJ....145....6J} and radio jets that extend over 43~arcmin across the sky. Assuming a distance to the galaxy of 28.5~Mpc its size corresponds to 355~kpc. To study the distribution of cold neutral hydrogen in and around the galaxy \ngc\ we used \ac{HIPASS} data \citep{2001MNRAS.322..486B} which did not show  H\,{\sc i} emission nor absorption associated with \ngc.

In this section, we present our observational results, and analyse the properties of the jet and its environment.
\subsection{Jet structure}

Fig.~\ref{figure:multi_Freq_ngc2663} shows detection of \ngc\ with \ac{ASKAP}, \ac{MWA} and \ac{ATCA} at three different frequencies, $1520$~MHz, $200$~MHz and $5500$~MHz respectively. The southern jet has an initial flare as it leaves its host galaxy, at $\sim 60$~arcsec from the core. The radio jets of \ngc\ have an opening angle of $\approx ~30^\circ$ in the innermost region. After this initial increase in width, the jet maintains a roughly constant width to the outer regions of the jet. Similar jet expansion profiles of 15 other \ac{AGN} jets have been reported by \citet{2021A&A...647A..67B}. 

\begin{figure*}
\centering
\includegraphics[width=\textwidth,trim = 120 0 180 0, clip]{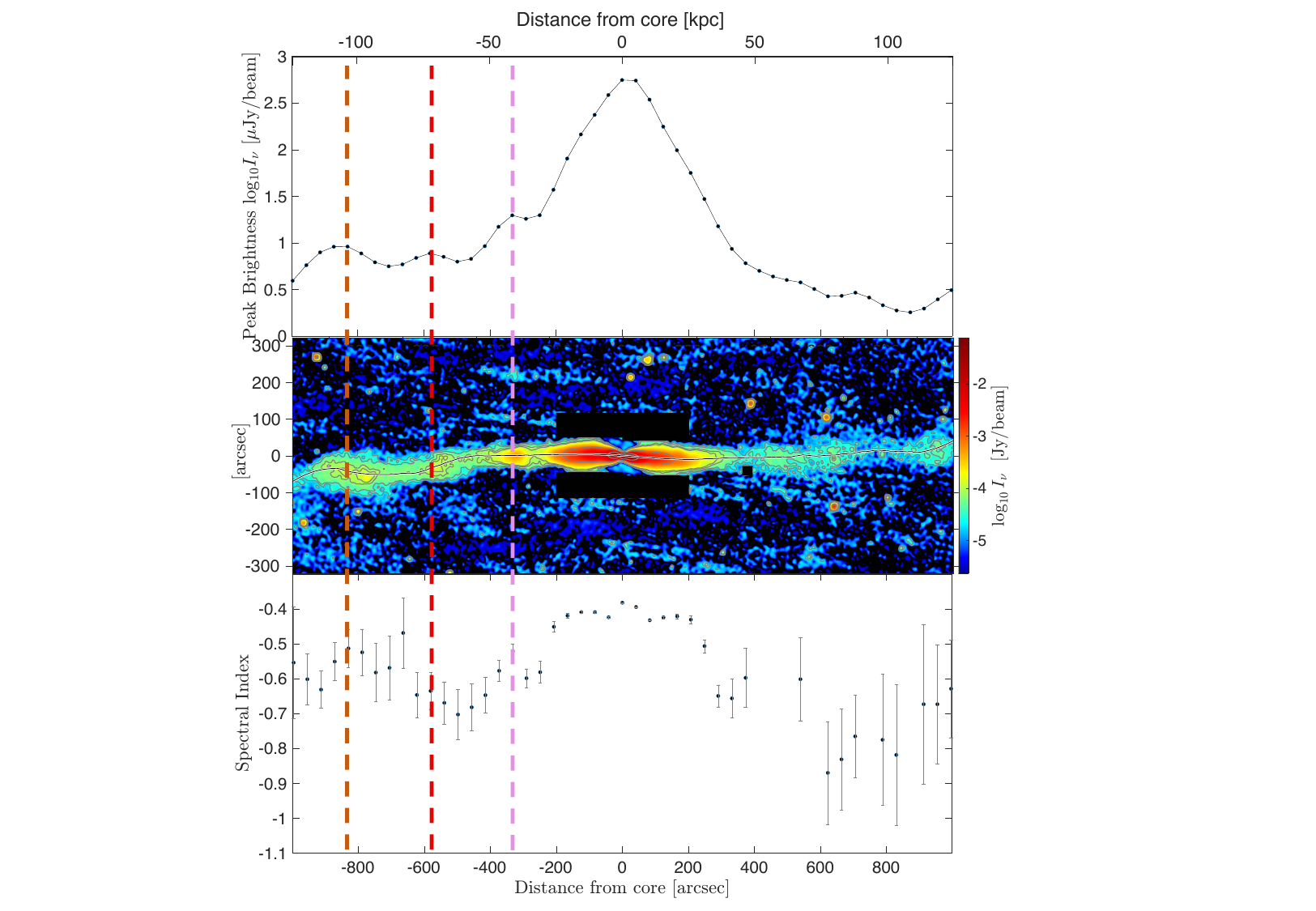}
\caption{
Top: Peak brightness, measured from ASKAP data, as a function of the distance from the core. Middle: ASKAP image of \ngc\ with the southern jet on the left side. Side lobes above and below the centre of the jet, and a point source, have been masked. The centre black line is the ridge-line of the jet, found as a result of the Gaussian fitting. Bottom: Spectral index calculated after convolving image from \ac{ASKAP} 1500~MHz to the resolution of \ac{MWA} 200~MHz image with beam size of $64.4\times49.0$~arcseconds.}
\label{fig_spec_index}
\end{figure*}

In the southern jet  (left in Figure \ref{fig_spec_index}), we see at least one and plausibly three regions (knots) in which the jet increases significantly in brightness. These are located at $\sim$295, $\sim$530 and $\sim$690~arcsec (41, 73 and 95~kpc) from the core. They resemble similar features as in other \ac{AGN} jets that are identified as (re)collimation shocks on pc scales. Recollimation is characterised by a simultaneous narrowing and brightening of the flow, as the flow lines converge, followed by a broadening and fading of the jet. We will examine this scenario further in the next section.


\begin{figure*}
    \centering
        \includegraphics[width=\textwidth,trim = 0 70 0 60, clip]{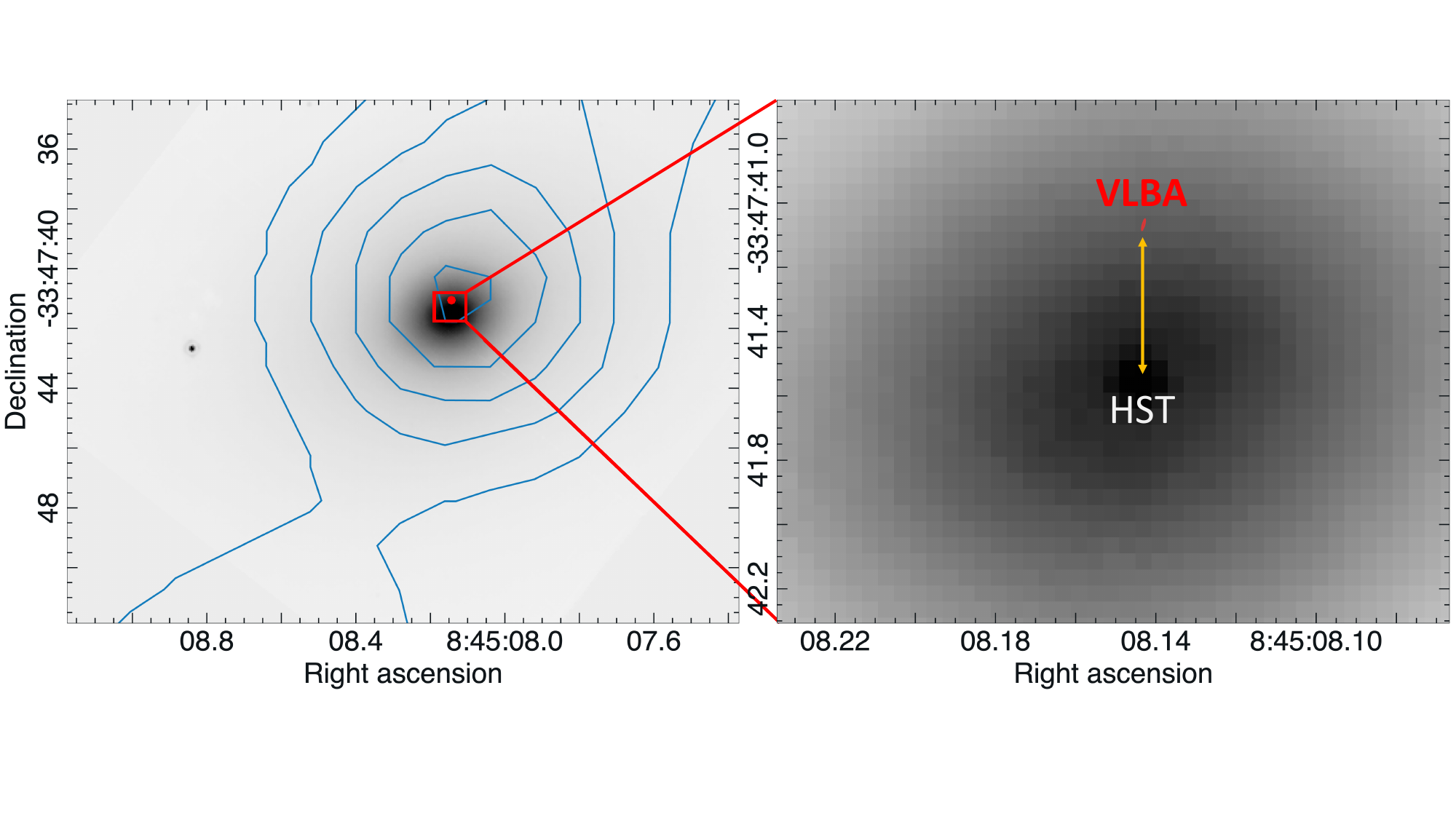}
        \caption{\ac{HST} image (grey scale) of \ngc\ central part (core) overlaid with the \ac{ASKAP} contours (blue; 10, 30, 50, 70, 90 and 110 mJy~beam$^{-1}$). The red dot indicates the position of the \ngc\ \ac{VLBA} core. The resolution of the \ac{HST} image is 50~mas while the \ac{VLBA} 8646~MHz image beam size is 1.13$\times$0.74~mas. }
\label{fig:HST}
\end{figure*}

Fig.~\ref{fig:HST} shows a positional offset between the \ngc\ \ac{HST} centre (at RA(J2000)=8$^{\rm h}$45$^{\rm m}$8.141$^{\rm s}$ and Dec(J2000)=33\D47\arcmin41.54\arcsec) and \ac{VLBI} (RA(J2000)=8$^{\rm h}$45$^{\rm m}$8.144$^{\rm s}$ and Dec(J2000)=33\D47\arcmin41.07\arcsec; $\sigma=1$~mas). This 0.47$\pm$0.3~arcsec offset equates to $\sim$70~pc at the distance of \ngc. At the time of the observations (2007) the \ac{HST} pointing accuracy was 0.3~arcsec\footnote{\nolinkurl{https://hst-docs.stsci.edu/drizzpac/chapter-24-astrometric-informationin-the-header/4-5-absolute-astrometry}} which indicates that this discrepancy might largely be due to the telescope astrometric error. However, this shift in position might also indicate that the central \ac{SMBH} is offset from the optical/IR centre of \ngc. This is not unusual: \citet{2017ApJ...835L..30M} present astrometric evidence for a population of dislodged \ac{AGN}. The most massive elliptical galaxies (such as \ngc) have low-density cores that differ significantly from the high-density centers of less massive ellipticals and the bulges of disk galaxies \citep{2014ApJ...782...39T}. These low-density cores are most likely the result of \ac{SMBH} binary mergers, which depopulate galaxy centers by gravitationally sling-shotting central stars outward. Such binaries naturally form in mergers of luminous galaxies \citep{2014ApJ...782...39T}. Alternatively, this offset might be due to the VLBA detecting an inner knot of the radio jet rather than the galaxy nucleus.

\subsection{Polarisation and rotation measure}

We made images of I, Q  and U polarisation and apparent magnetic field direction, the latter by correcting the observed angles for Faraday rotation determined from the two frequencies, when the RM could be reliably determined.

\begin{figure*}
\centering
\includegraphics[width=\textwidth]{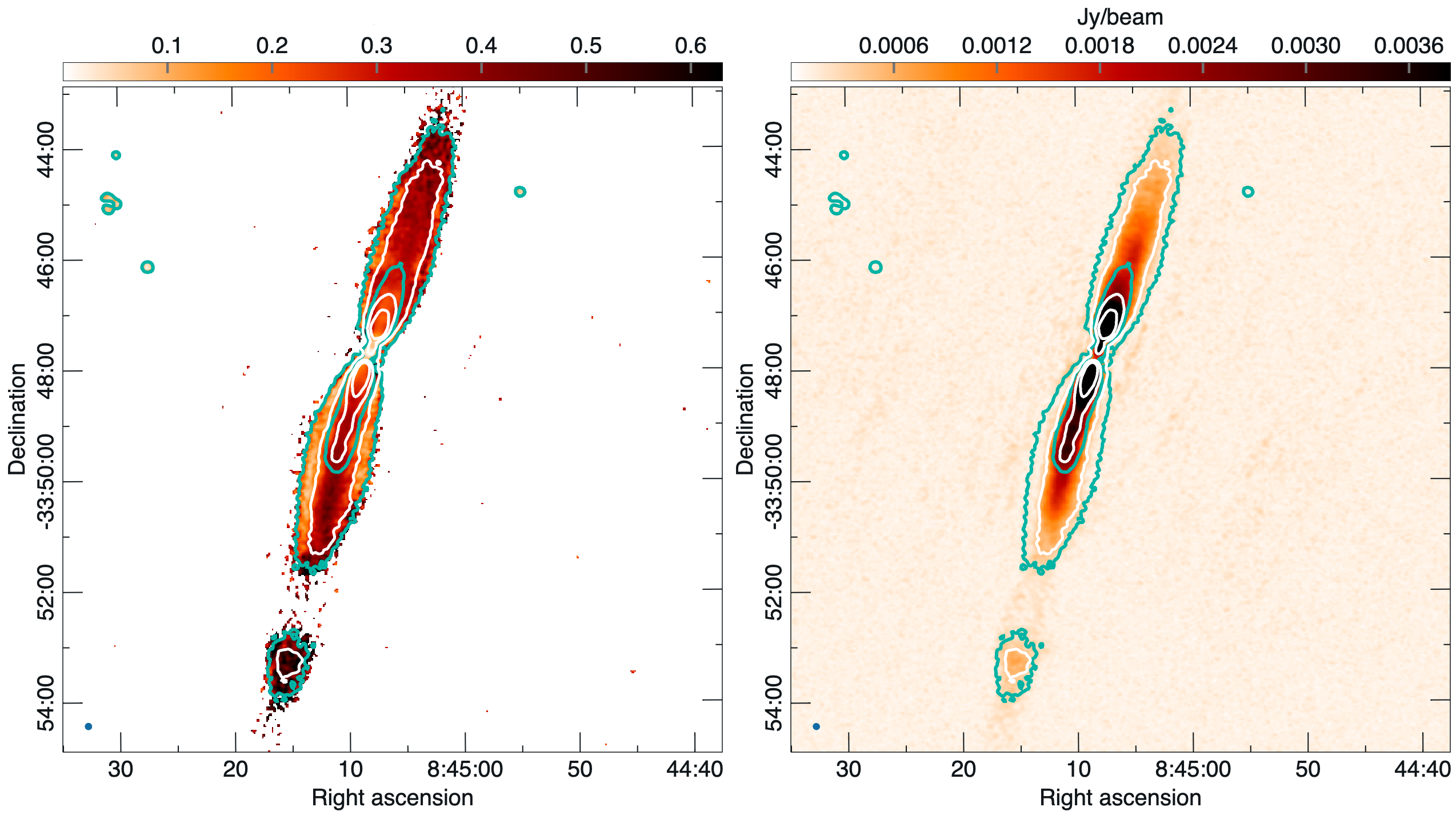}
\caption{ \emph{Left:} Fractional polarisation of \ngc\ obtained with \ac{ASKAP}. The highest fractional polarisation occurs along the ridge line and drops towards the edges. \emph{Right:} Polarization intensity obtained with \ac{ASKAP}. Teal contours represent total intensity at $5\sigma$ and $50\sigma$ with $\sigma = 48$~$\mu$Jy~beam$^-1$, while white contours are taken from the polarization intensity map at ($5,30,70)\times \sigma$, with $\sigma = 93$~$\mu$Jy~beam$^-1$.
The beam size is shown in the bottom left corner as a blue circle ($6 \times 6$~arcsec).}
\label{fig_polarisation}
\end{figure*}

Fig.~\ref{fig_polarisation} shows dominant linear polarization along the jets. It shows an interesting feature where fractional polarization is highest along the ridge line and drops towards the edge of the jet. In the southern jet around $\sim40$~kpc fractional polarisation significantly increases in the region where the knot appears to be.

\begin{figure*}
\centering
\includegraphics[width=\textwidth]{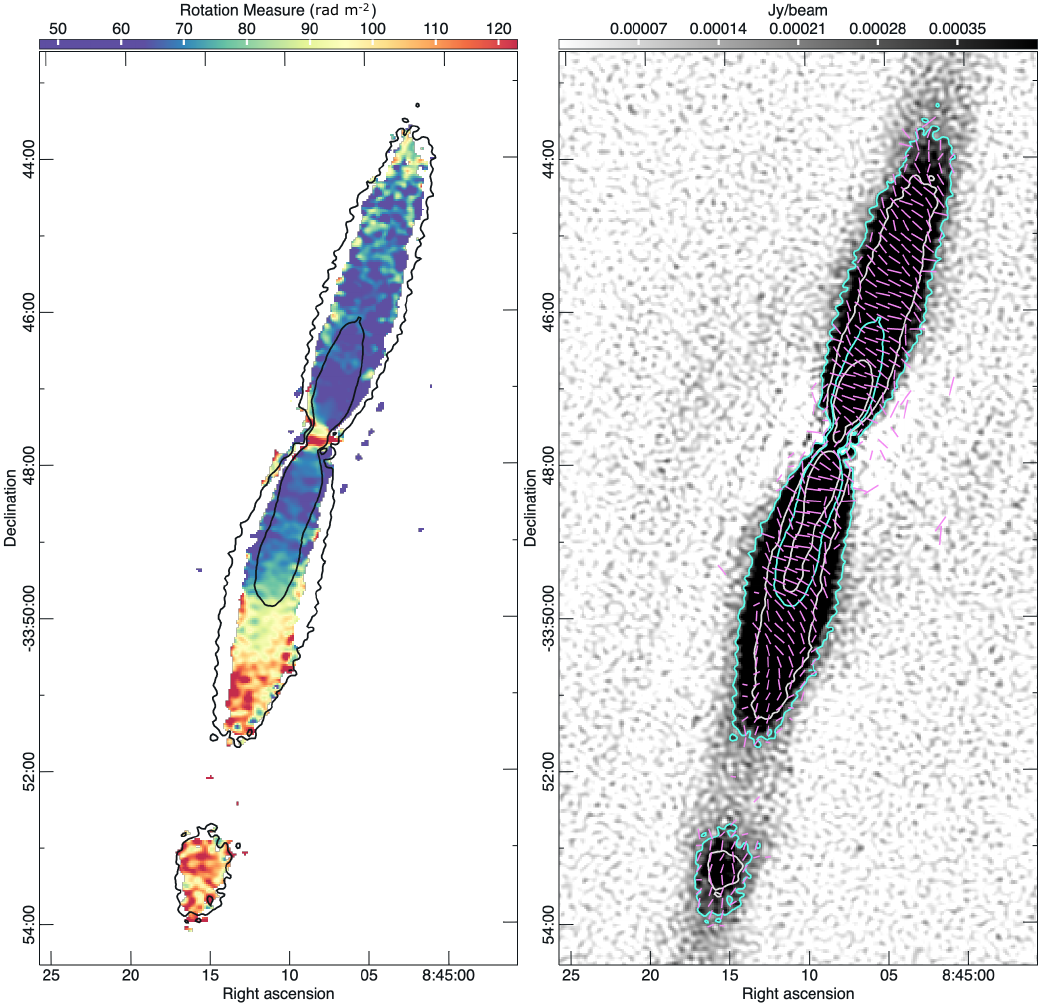}
\caption{
    \emph{Left:} RM colour map of \ngc\ corresponding to peak polarisation intensity after RM Synthesis overlaid with black contours which are taken from the total intensity map at ($5,50)\times \sigma$, with $\sigma = 48$~$\mu$Jy~beam$^{-1}$. \emph{Right:} Total intensity image overlaid with magnetic field vectors derived after RM correction (magenta lines), contours are identical to the ones in Fig.~\ref{fig_polarisation}, beam size is ($6 \times 6$~arcsec).
 }
\label{fig:pol_ASKAP}
\end{figure*}

Fig.~\ref{fig:pol_ASKAP} shows a map of RM and total intensity image overlaid with contours of the Stokes I image and magnetic field vectors (magenta lines) which are predominantly oriented transverse to the jet axis, which is expected in \FRI\ jets \citep{1994RPPh...57..325K}. However, the magnetic field vectors in the knot are aligned with the jet direction.
In the northern jet and the inner portion of the southern jet we calculate similar narrow distributions of RM with a mean of $\sim 66$~rad~m$^{-2}$, at approximately 2 arcmin from the core in the southern jet the distribution of RM broadens with a mean of $\sim105$~rad~m$^{-2}$. This suggests that radiation originating from the southern jet passes through more surrounding material which suggests that \ngc\ is slightly inclined to the plane of the sky, with the northern jet pointing towards us \citep{1988Natur.331..147G}.
Using the CIRADA cutout server\footnote{\nolinkurl{http://cutouts.cirada.ca/rmcutout/}} we obtained the expected Galactic foreground rotation measure at the position of \ngc. The northern side shows $ 67\pm7$~rad m$^{-2}$, while the southern side of the \ngc\ has Galactic foreground emission of $70 \pm 7$ rad m$^{-2}$ \citep{2022A&A...657A..43H}.

\subsection{Spectral index}

The spectral index of \ngc, defined using $F_\nu \propto \nu^{\alpha}$, was calculated using images obtained with the \ac{MWA} at 200~MHz and \ac{ASKAP} at 1500~MHz convolved to the same resolution of $64.4\times49.0$~arcsec. We performed 1D Gaussian fits to cuts perpendicular to the jet axes in the two images and used their peak brightnesses to get the indices. 
In the bottom panel of Fig.~\ref{fig_spec_index} we show the distribution of the spectral index along both jets where the southern jet is on the left. The \ngc\ jets have a spectral index of $\alpha=-0.42$ for up to $\sim$200~arcsec ($\sim$40~kpc) on each side of the core, far beyond of what would be expected from a compact core alone.
It is flatter than in other \ac{AGN} jets on kpc scales (see: \citealp{1997ApJ...488..146K}, \citealp{2008ASPC..386..110L}, \citealp{2008MNRAS.386..657L},  \citealp{2020MNRAS.495..143C}). We note that spectra of the base of the jets using higher resolution, higher frequency observations  from \ac{ATCA}  (1384~MHz and 2368~MHz) show spectral indices of $\sim$ --0.7 in this region. Observations at other low frequencies would be needed to confirm the apparent spectral curvature.

Further along the southern jet in a region from 200 to 400~arcsec (40-80~kpc; marked with the pink dashed line on Fig.~\ref{fig_spec_index}) the spectrum sharply steepens to $\alpha=-0.62$ before flattening to about $\alpha=-0.5$ at the first knot. We observe a similar effect in regions 400 to 600~arcsec (red dashed line in Fig.~\ref{fig_spec_index}), 600 to 800~arcsec (dark red dashed line in Fig.~\ref{fig_spec_index}) and even beyond 800~arcsec from the core of \ngc. This may indicate that the jet has been re-energised at the knots. In the north, the spectral index steepens to $\alpha \approx -0.9$, with the steepening leveling off further out. These steepening patterns likely represent a combination of spectral losses and changing magnetic fields. However, with the current data, it is not possible to tell whether the leveling off and flattening requires particle acceleration or can simply be explained by changes in the magnetic field strength.

\subsection{Host galaxy: WISE}

\begin{figure*}
    \centering
    \includegraphics[width= \columnwidth]{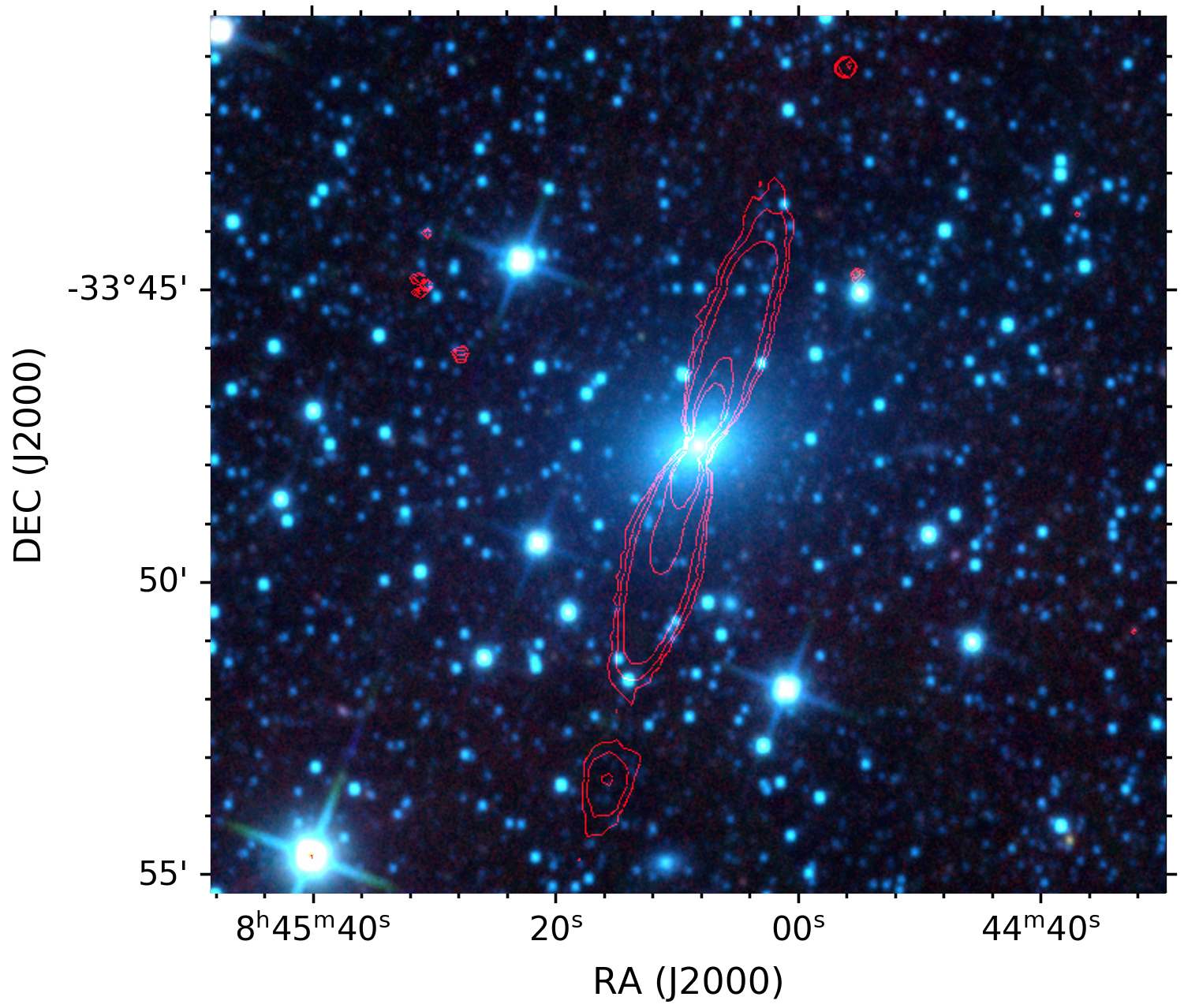}
    \includegraphics[width=\columnwidth]{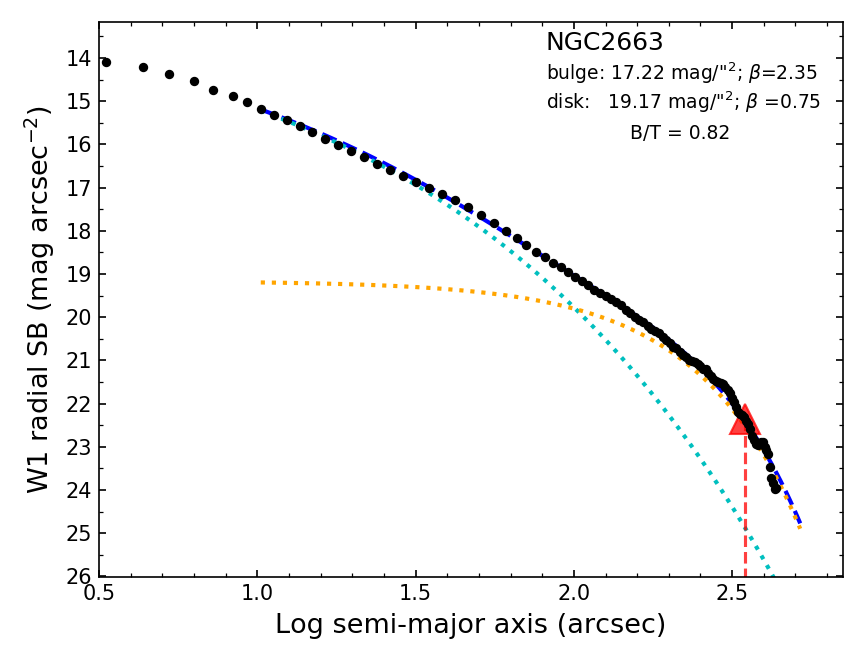}
    \caption{\emph{Left:} color-composite of the WISE image of \ngc\ using a combined W1 (3.4$\mu$m; blue), W2 (4.6$\mu$m;green) and W3 (12$\mu$m;red) imaging.  The radio continuum from ASKAP (1520 MHz) is shown with red contours ranging from 0.15, 0.25, 0.5, 2.5, 5~mJy~beam$^{-1}$.  
    \emph{Right:} W1 azimuthally averaged radial surface brightness distribution (black point), overlaid with a double-sersic bulge (blue) + disk (orange) model \citep{2019ApJS..245...25J}.  The 1-$\sigma$ isophotal radius is denoted by the red triangle. The bulge-to-total ratio is 0.82, indicating a bulge-dominated host galaxy.} 
    \label{fig:SED}
\end{figure*}

Fig.~\ref{fig:SED} (left) shows the mid-infrared view from \ac{WISE} using the stellar-sensitive bands at 3.4$\mu$m (W1) and 4.6$\mu$m (W2), and the ISM sensitive band at 12$\mu$m (W3).
They reveal that \ngc\ is a typical red-and-dead elliptical galaxy. It shows little evidence of star formation or a gas disk. It is nearly dust-free. The central region is detected in the $12$ and $22~\mu$m bands. This emission arises from the old stellar population, and likely, from warm/hot dust in the \ac{AGN} \citep{2017ApJ...850...68C}, which is consistent with low-power \FRI\ AGN being radiatively inefficient \citep{1979MNRAS.188..111H}. Infrared observations from \ac{WISE} show no sign of dust associated with the \ac{SMBH} accretion disk. 

The host galaxy is very bright and luminous, and dominates any radiation from the \ac{AGN}. Fig.~\ref{fig:SED} (right) shows the axi-symmetric radial surface brightness profile of stellar light, from which we see that \ngc's elliptical shape has a bulge-to-total ratio of 0.82, and hence is dominated by the bulge/spheroidal stellar population \citep{2013AJ....145....6J,2019ApJS..245...25J}.

\subsection{X-ray emission}
\label{XraySubSection}
An analysis of the X-ray emission from \ngc\ as observed with three different instruments shows complementary results. The environment surrounding \ngc\ is rarefied: there is no diffuse emission except in the vicinity of the core. No X-ray emission is detected from the jet. This is certainly unusual for AGN, and is discussed more in the following sections.

\subsubsection{Chandra X-Ray Observatory}
 \label{chandra}

\begin{figure}
    \centering
    \includegraphics[width=\columnwidth,trim = 100 15 150 0, clip]{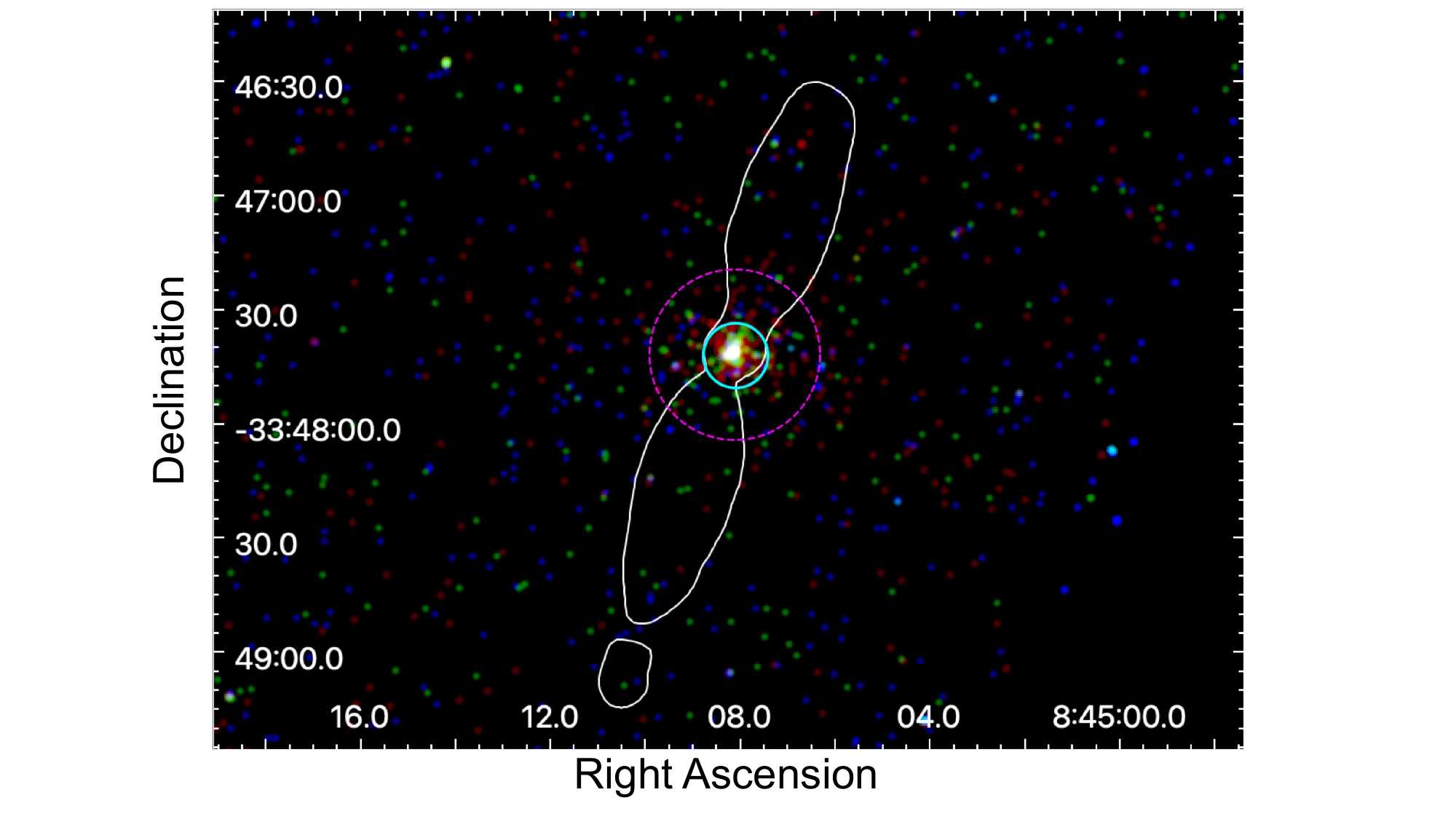}
    \caption{\cxo\ three-color X-ray image of \ngc\ with \ac{ASKAP} contours overlaid. The red, green and blue color scale corresponds to soft X-ray (0.5~keV to 1.2~keV), medium X-ray (1.2~keV to 2.0~keV) and hard X-ray (2.0~keV to 7.0~keV) emission, respectively. Emission is detected in the close vicinity of the black hole and dominant in soft and medium bands, 0.5-2~keV.
    White contours are obtained from ASKAP total intensity image at $0.04$~mJy, the cyan circle represent source extraction region of $8.5$~arcsec radius, while the magenta annulus between $8.5$~arcsec and $22.5$~arcsec represents background extraction region.}
    \label{fig_chandra}
\end{figure}

A \cxo\ image of \ngc\ is presented in Fig.~\ref{fig_chandra} with \ac{ASKAP} radio contours. The three-color image indicates that the central source produces a mixture of emission from the soft and medium energy bands (corresponding to energies of 0.5~keV to 1.2~keV and 1.2~keV to 2.0~keV, respectively). The emission appears to be elongated in the same direction as the radio jets.

A \cxo\ spectrum corresponding to the central region of \ngc\ was extracted using a circular region centred at RA~(J2000.0)=08$^h$~45$^m$~08.1$^s$ and Dec.~(J2000.0)=$-$33\D\ 47\arcmin\ 42.1\arcsec\ with radius of $8.5$~arcsec. A background spectrum was extracted from an annular region centred on the source region: the radius of this background extraction region was 14 arcseconds. The spectrum was fitted over the energy range from 0.7~keV to 3.0~keV with a combined model including the thermal component {\sc mekal} which describes an emission spectrum from a hot diffuse gas (\citet{1985A&AS...62..197M,1986A&AS...65..511M}; \citet{kaastra1992x}; \citet{1995ApJ...438L.115L}) and a power law  component.
The combined model was multiplied by the T\"{u}bingen-Boulder absorption model TBABS with elemental abundances as tabulated by \citet{2000ApJ...542..914W}) fixed to the Galactic column density $N_{H} = 2.35\times 10^{21}$~cm$^{-2}$ \citep{2005A&A...440..775K}. In addition, for the fit parameters of the {\sc mekal} model, the elemental abundances were frozen to solar values and the redshift was frozen to our adopted redshift value for NGC 2663 (that is, $z=0.00701$). Using this combined model, a statistically acceptable fit was obtained (C-Statistic of 8.22 for 9 degrees of freedom). The fitted value of the temperature of the {\sc mekal} component was $kT = 0.51^{+0.16}_{-0.20}$~keV and the fitted value of the photon index of the power law component was $\Gamma =2.36^{+0.77}_{-0.88}$ (quoted error bounds correspond to 90~per~cent confidence limits).

\begin{figure}
    \centering
        \includegraphics[width=\columnwidth,trim = 40 0 138 0, clip]{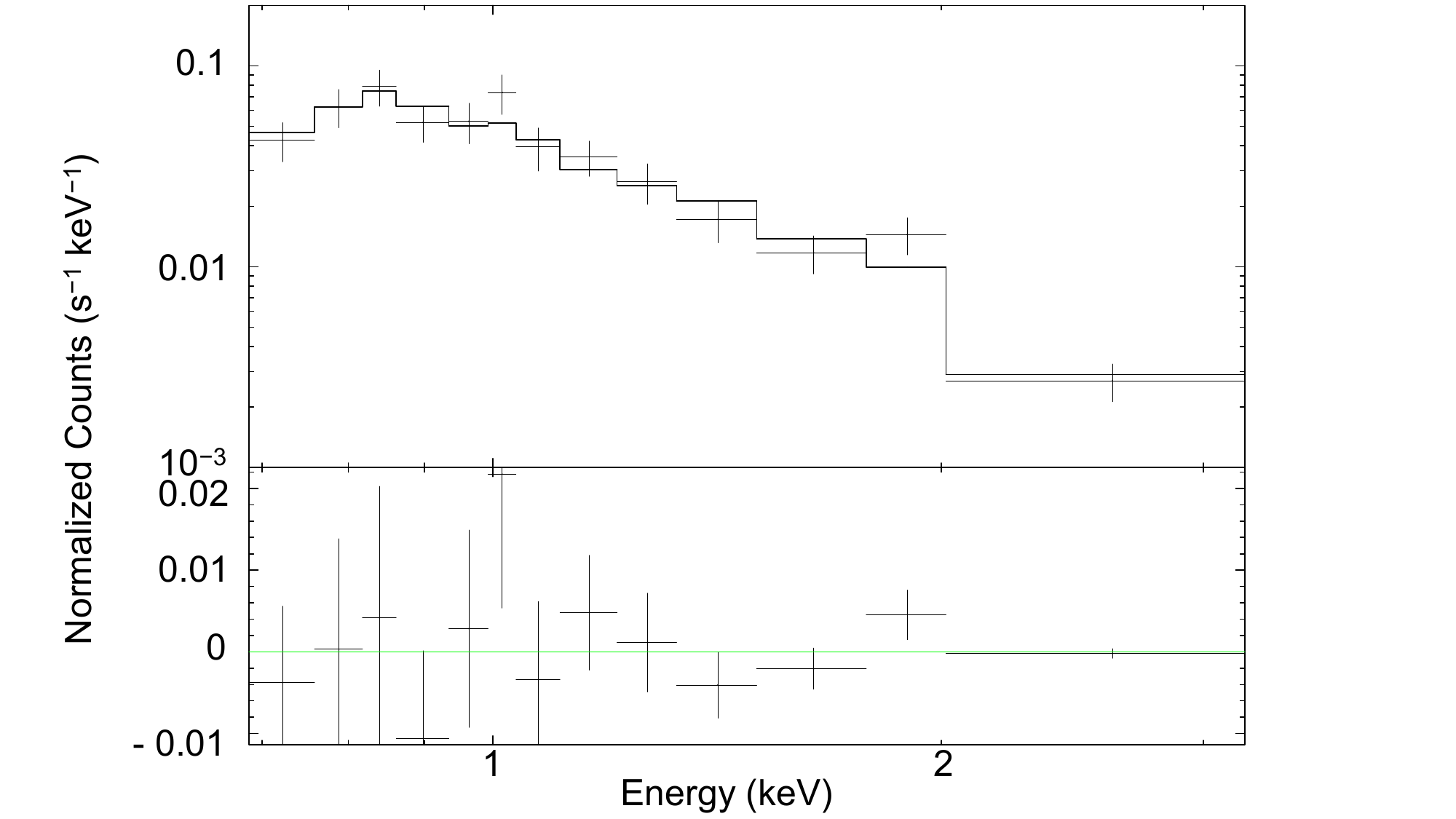}
        \caption{\cxo\ spectrum of \ngc\ as fit with the TBABS $\times$ ({\sc mekal} + Power Law).}
\label{fig_chandra_spectrum}
\end{figure}

The extracted \cxo\ spectrum as fit with this combined model is presented in Fig.~\ref{fig_chandra_spectrum}. 
The measured unabsorbed flux of this source over the energy range given above is $2.77\times10^{-16}$~W~m$^{-2}$. At the adopted distance to this source $z = 0.00701$ and assuming a Hubble's Constant value of 70 km/s/Mpc, this unabsorbed flux corresponds to an unabsorbed luminosity of $2.91\times10^33$~W.

We have also conducted a timing analysis of the X-ray emission from this region using the CIAO tool {\it glvary} to search for time variability in the X-ray emission from this source during the observation. This tool is based on the Gregory-Loredo method for identifying variability in the X-ray emission from sources \citep{1992ApJ...398..146G}. Analysis with this tool indicated no evidence for time-variable X-ray emission.

\subsubsection{X-ray \Swift}
\label{swift_res}

The X-ray spectrum from the \Swift\ X-Ray Telescope was fitted with a single power law model with the absorption at $z=0$ fixed to the Galactic column density $N_H = 2.35\times 10^{21}$~cm$^{-2}$ \citep{2005A&A...440..775K} resulting in a photon index $\Gamma = 2.37^{+2.23}_{-1.46}$. The observed flux in the 0.3--10~keV band is 2.7$^{+2.0}_{-0.9}\times 10^{-16}$~W~m$^{-2}$ which is $5.0\times 10^{-16}$~W~m$^{-2}$ corrected for Galactic absorption. Given the \ngc\ distance of 28.5~Mpc the luminosity of the source is $5\times 10^{33}$~W.

\subsubsection{UV \Swift}
The magnitudes in u and uvw1 in the \Swift\ UVOT in the first observation were $16.41\pm 0.05$ (14.71) mag and $17.58\pm 0.10$ (15.45) mag, respectively with the values in parenthesis denoting the magnitudes corrected for Galactic reddening. The measurements of the second observation do not suggest any variability in the UV between the two observations. 

\subsubsection{eROSITA}
\label{erosita}

\begin{figure}
	\centering
	\includegraphics[width=\columnwidth,trim = 30 20 35 25, clip]{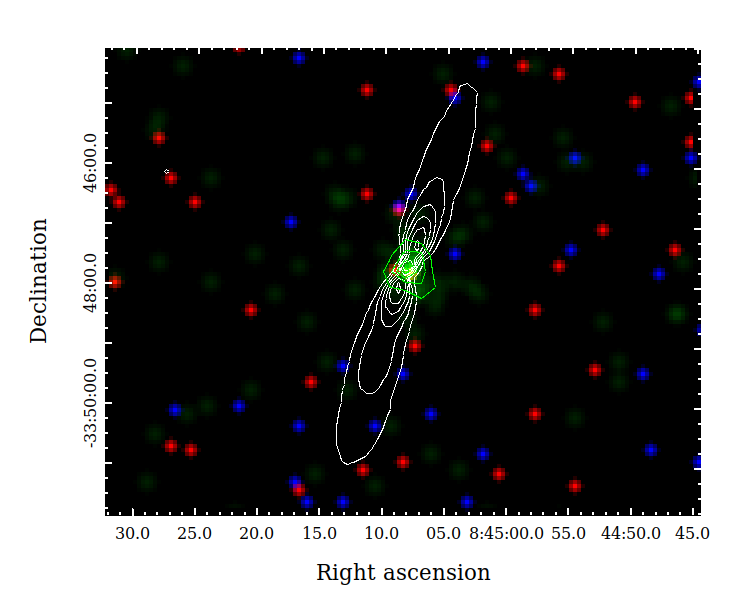}
	\caption{Smoothed eROSITA X-ray image of \ngc\  where red corresponds to the $0.2-0.6~{\rm keV}$ band, green is the $0.6-2.3~{\rm keV}$ band, and blue is the $2.3-5~{\rm keV}$ band. Contours are also over-plotted, the X-ray contours from the eROSITA "Green" image with energies $0.6-2.3$~keV shown in green and radio contours from ASKAP at $1,3,5,8,15,27$~mJy~beam$^{-1}$ shown in white.}
	\label{fig:e_rgb}
\end{figure}

Fig.~\ref{fig:e_rgb} shows the smoothed three-colour X-ray image from eRASS1, where red corresponds to the $0.2-0.6~{\rm keV}$ band, green is the $0.6-2.3~{\rm keV}$ band, and blue is the $2.3-5~{\rm keV}$ band. The blue and red (soft and hard) events in the image correspond to individual photons and are therefore associated with the X-ray background and not to real detected sources. The source is centred in the image, and is most significantly detected in the $0.6-2.3~{\rm keV}$ band and thus shown primarily in green. Over-plotted on the image are the X-ray contours from eROSITA (shown in green) and radio contours (shown in white). The core of \ngc\ is detected with eROSITA, and the innermost radio position is matched with X-ray contours. There is no evidence for X-ray emission from \ngc\ beyond the galaxy core, nor is there X-ray emission
from a surrounding group medium.

\begin{figure}
	\centering
	\includegraphics[width=\columnwidth]{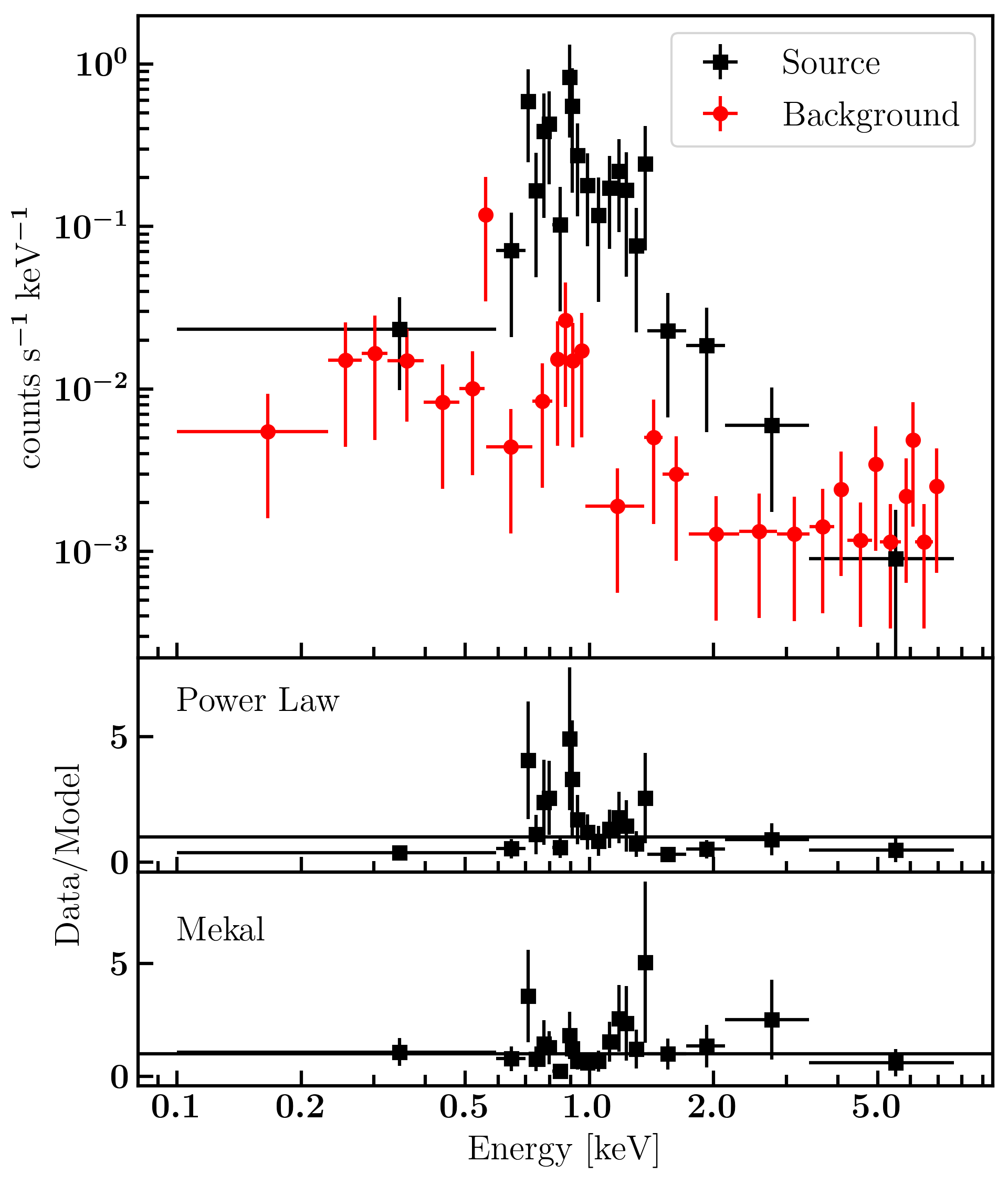}
	\caption{Top panel: source (black squares) and background (red circles) eROSITA spectra for \ngc\ . Spectra have been re-binned in XSPEC for clarity. The source is brightest from $\sim0.6-3~{\rm keV}$, and is background dominated at lower and higher energies. Middle panel: residuals (data/model) from an absorbed power law fit. Bottom panel: residuals (data/model) from the absorbed {\sc mekal} model.}
	\label{fig:e_dat}
\end{figure}

The top panel of Fig.~\ref{fig:e_dat} shows the source (black squares) and background (red circles) spectra acquired by eROSITA. Spectra have been re-binned in XSPEC for clarity. The source appears background dominated above $\sim3.5~{\rm keV}$, as well as below $\sim0.6~{\rm keV}$; likely due to the high level of Galactic absorption (see \ref{chandra}). The middle panel shows the residuals (data/model) from a power law modified by Galactic absorption. The data are over-fit by this model, with a C-statistic of 59 for 109 degrees of freedom. A best-fit photon index of $\Gamma = 2.67_{-0.49}^{+0.66}$ is measured, in agreement with both \cxo\ and \Swift. This corresponds to a $0.3-10~{\rm keV}$ luminosity of 1.2$\times$10$^{34}$~W, a factor of $\sim2$ higher than the luminosity from the \Swift\ observation and about 5 times higher than the one detected with Chandra. This value is slightly lower than what is likely to be observed in a typical \ac{AGN}. Furthermore, an examination of the $0.5-2.3~{\rm keV}$ light curve shows no evidence for variability over the course of the observation.

The eROSITA analysis is highly sensitive to the soft band, seems to prefer an ionised plasma origin for the emission rather than a purely jet-based origin. The ionized plasma may only be due to the host galaxy plasma, or star formation, or a combination of both. 
From the power law fit, clear positive and negative residuals remain across the spectrum.  To model this, the power law component is replaced with {\sc mekal}, to model emission from hot, diffuse gas \citep{1995ApJ...438L.115L}. The residuals are shown in the bottom panel of Fig.~\ref{fig:e_dat}. The fit improves by a $\Delta C$ of 6 for no additional degrees of freedom, and many of the residuals improve. Furthermore, the best fit plasma temperature of $kT = 0.68_{-0.10}^{+0.16}~{\rm keV}$ and $0.5-2~{\rm keV}$ luminosity of 5.5$\times$10$^{33}$~W are in agreement with those seen in hot, diffuse gas of LLAGN \citep{2003MNRAS.343.1181F}.

Using the same data we also fitted the combined model, {\sc mekal} component and an absorbed power law. From this analysis we got a temperature of $T = 0.56$~keV and a photon index of $\Gamma = 2.04$. The Bayesian evidence for the ({\sc mekal} + power law) model is $-431$  and $-418$ for {\sc mekal} alone, where the lower number indicates a better fit. The reason {\sc mekal} model is a better fit is probably because the {\sc mekal}l component dominates the soft spectral shape where the eROSITA is highly sensitive, while the combined ({\sc mekal}+power law) model is too complex for the given data. 
The slight difference between \cxo\ and eROSITA might also indicate that the \ngc\ varied between the two observations.

\section{Discussion}
\label{sect:DC}
The multi-frequency appearance of \ngc\ raises a number of questions about the physical causes that determine the size and shape of its \ac{AGN} jet. Why is it so large, and why does it seem to recollimate? What combination of nature (the \ac{AGN}) and nurture (the galactic and intergalactic environment) best explains its properties? It is quite important for our understanding of jet physics to establish whether recollimation --- the physical, rather than merely apparent, narrowing of the jet --- is taking place. \ngc\  provides us one of the best current opportunities to do that, so we examine this possibility in detail.
We will first present the case for recollimation, and then consider some alternative mechanisms.

\begin{figure*}
    \centering
    \includegraphics[width=\textwidth,trim = 210 0 210 0, clip]{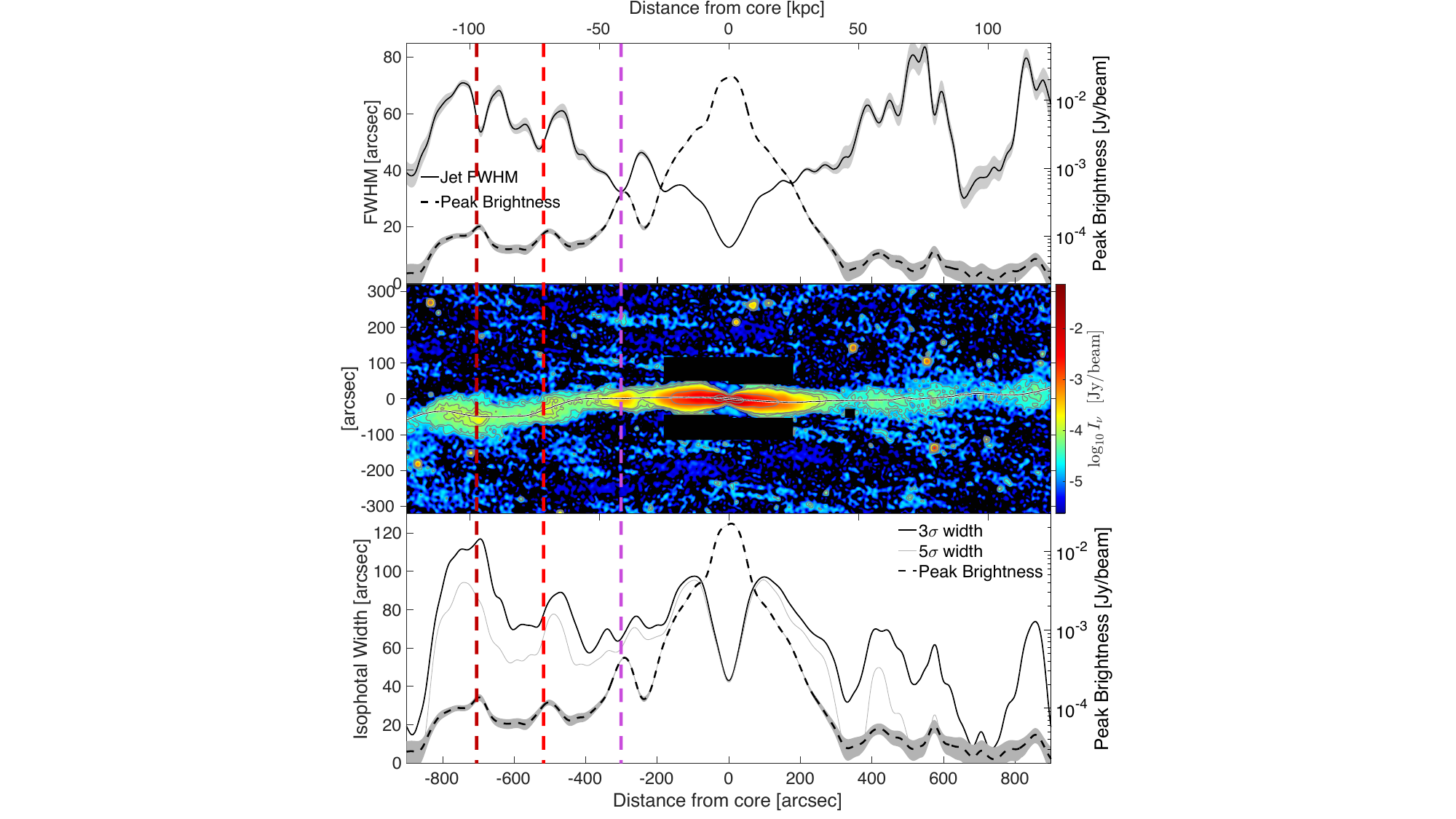}
    \caption{\emph{Top:} \ac{FWHM} jet width (solid) and peak brightness (dashed), measured from the \ac{ASKAP} data, as a function of the distance from the core. These quantities are calculated by least-squares fitting a Gaussian to each vertical column of pixels. This gives the location of the ridge-line of the jet, the peak brightness along the ridge-line, and the \ac{FWHM} width of the jet.
    \emph{Middle:} ASKAP image of \ngc\ with southern jet on the left side. The two lowest contours in the bottom plot are $3\sigma$ and $5\sigma$ above the background noise. Side lobes above and below the centre of the jet, and a point source, have been masked. The solid black line is the ridgeline of the jet, found as part of the Gaussian fitting. 
    \emph{Bottom:} Isophotal (constant brightness) width of the jet, measured from \ac{ASKAP} observations, of the $3 \sigma$ (black line) and $5\sigma$ (grey) contours, where $\sigma$ is the background noise. The dashed line shows the peak brightness of the ridge-line of the jet.}
    \label{plot:jetwidth}
\end{figure*}

\subsection{Bright knots: recollimation?}

As some of the terms relevant for this discussion are not used consistently in the literature, we start with some definitions. We assume that, following an initial magnetically-dominated acceleration phase on parsec scales, a conical, hydrodynamic jet emerges with a certain opening half-angle $\theta$ \citep[as in, for example, M\,87;][]{2014ApJ...781L...2A}.

\emph{Collimation} occurs if, for internal or external reasons, the jet becomes cylindrical, travelling with an approximately fixed width. Many (perhaps most) \ac{AGN} jets, after an initial widening stage, are observed to travel large distances without observably diverging \citep{2021ApJ...917...18C}, suggesting nearly parallel streamlines in the jet fluid.

\emph{Recollimation} occurs when an already collimated jet narrows (rather than merely straightens). This has elsewhere been called reconfinement or pinching. This may occur because of external, environmental forces on the jet, or because of internal forces or instabilities in the jet.

Fig.~\ref{plot:jetwidth} (top) shows the \ac{FWHM} width of the jet (black solid line) and the peak brightness (dashed line) of a cross-section, as a function of the distance from the core, for both jets of \ngc. These quantities are calculated by least-squares fitting a Gaussian to each vertical column of pixels. This gives the position of the ridge-line of the jet, the peak brightness along the ridge-line, and the \ac{FWHM} width of the jet. The fitting function also gives the errors in the top plot; in addition, the peak brightness error includes the contribution of background noise.

We see that at $\sim295$~arcsec ($\sim$43~kpc) from the core, the \ac{FWHM} of the southern jet simultaneously narrows while the jet brightens. However, lower brightness contours show a relatively constant width (Section~ \ref{sec:jet_structure}). Further brightening is seen at $\sim$530 and $\sim$690~arcsec ($\sim$73, $\sim$95~kpc), with marginal evidence for a narrowing of the \ac{FWHM} width of the jet at these positions. 

The environment of the jet plays a crucial role in its evolution.  The initially relativistic jet may entrain external material and slow down; this entrainment can be mediated by either Kelvin-Helmholtz instabilities (in the case of ambient gas) or stellar winds \citep{1994ApJ...422..542B,2021NewAR..9201610K,2006A&A...456..493P,2015MNRAS.447.1001W}. The final morphology of jet-inflated structures depends sensitively on this jet-environment interaction: jets which manage to stay relativistic form powerful \FRII\ edge-brightened structures, while the slower entrained jets form edge-darkened \FRI's, of either jetted or lobed morphology \citep{2018NatAs...2..273H}. 

To interpret the observed knots as recollimation shocks requires a combination of low ($\leq 0.01c$) jet velocity and poor environment ($n_{\rm x} \sim 0.001\,{\rm cm}^{-3}$).

\citet{2006MNRAS.368.1404A} and \citet{2012MNRAS.427.3196K} used a combination of analytical and numerical modelling to relate jet and environment properties to the final radio source morphology which is shown in \textbf{Equation \ref{eq:1}}.

\begin{equation}
\begin{aligned}
L_{1a} = 13 \, {\rm kpc} &\left( \frac{f_\theta}{1.9} \right) \left( \frac{v_{\rm j}}{0.1 c} \right)^{-1/2} \left( \frac{Q_{\rm jet}}{10^{36}\,{\rm W}} \right)^{1/2} \\
&\quad\left( \frac{n_{\rm x}}{0.001\,{\rm cm}^{-3}} \right)^{-1/2} \left( \frac{c_{\rm x}}{500\,{\rm km/s}} \right)^{-3/2}
\end{aligned}
\label{eq:1}
\end{equation}

$L_{1a}$ corresponds to the onset of jet recollimation, identified as the location where the jet sideways ram pressure and ambient pressure become equal.
The jet becomes collimated before its forward ram pressure drops to the ambient value; such a collimated jet can ``drill through'' the surrounding gas, inflating an FR-II structure. Conversely, if the jet runs out of forward thrust before recollimation is complete, it will be disrupted and form an \FRI\ structure.

Hence, a typical \FRI\ jet in a low density environment is expected to show signs of recollimation on scales of $\sim 10$ kiloparsecs. Richer environments, such as galaxy clusters, will yield collimation length scales shorter by a factor $n_x^{-1/2}$, corresponding to $\sim 1$~kpc in clusters. 

Whether the environment density of \ngc\ is sufficiently low for such large recollimation scales may be probed using Rotation Measure observations in the future.

\subsection{Bright knots: internal jet structure?}
\label{sec:jet_structure}
Could the bright knots in the jet of \ngc\ result from processes that are internal to the jet itself? In simulations and analytic models, jets can be surrounded by a cocoon or sheath of over-pressurised material, through which a narrower spine continues to travel \citep{2014MNRAS.437.3405L}.

If the jet has this kind of internal structure, the simultaneous brightening and narrowing of the knots can be explained without any evolution in the actual width of the jet. When the jet is relatively faint, the \ac{FWHM} is measuring the width of the (wider) sheath of the jet. But if the spine brightens (for some reason) in a certain region, the \ac{FWHM} measures the width of the (narrower) spine only. In this scenario, the bright knots in the jet of \ngc\ reveal the existence of a narrow spine, rather than resulting from the constriction of the jet by its environment.

Observational evidence relevant to this scenario comes from considering the isophotal width of the jet, looking for multiple components in cross-sectional slices, and the linear polarisation structure.

Fig.~\ref{plot:jetwidth} shows the isophotal width. The middle panel shows the \ac{ASKAP} brightness image of \ngc, where the two lowest contours in the bottom plot are 3$\sigma$ and 5$\sigma$ above the background noise. The 3$\sigma$ and 5$\sigma$ brightness contours show a remarkably constant width between $\sim 150$ and $600~$arcsec ($\sim 20-80~$kpc), seemingly oblivious to the rapid, four-fold brightening and fading of the spine of the jet. This can be seen in the bottom panel, which shows the isophotal width of the 3$\sigma$ (solid black line) and 5$\sigma$ (solid grey line) \ac{ASKAP} brightness contours. This striking constancy of the isophotal width is consistent with a component of the jet that is decoupled from the cause of the brightening of the knot. However, the evolution of the brightness of the jet --- with changing density, internal and external pressure, magnetic field, temperature, etc. --- makes a contour of fixed brightness difficult to interpret.

\begin{figure}
    \centering
    \includegraphics[width=0.5\textwidth]{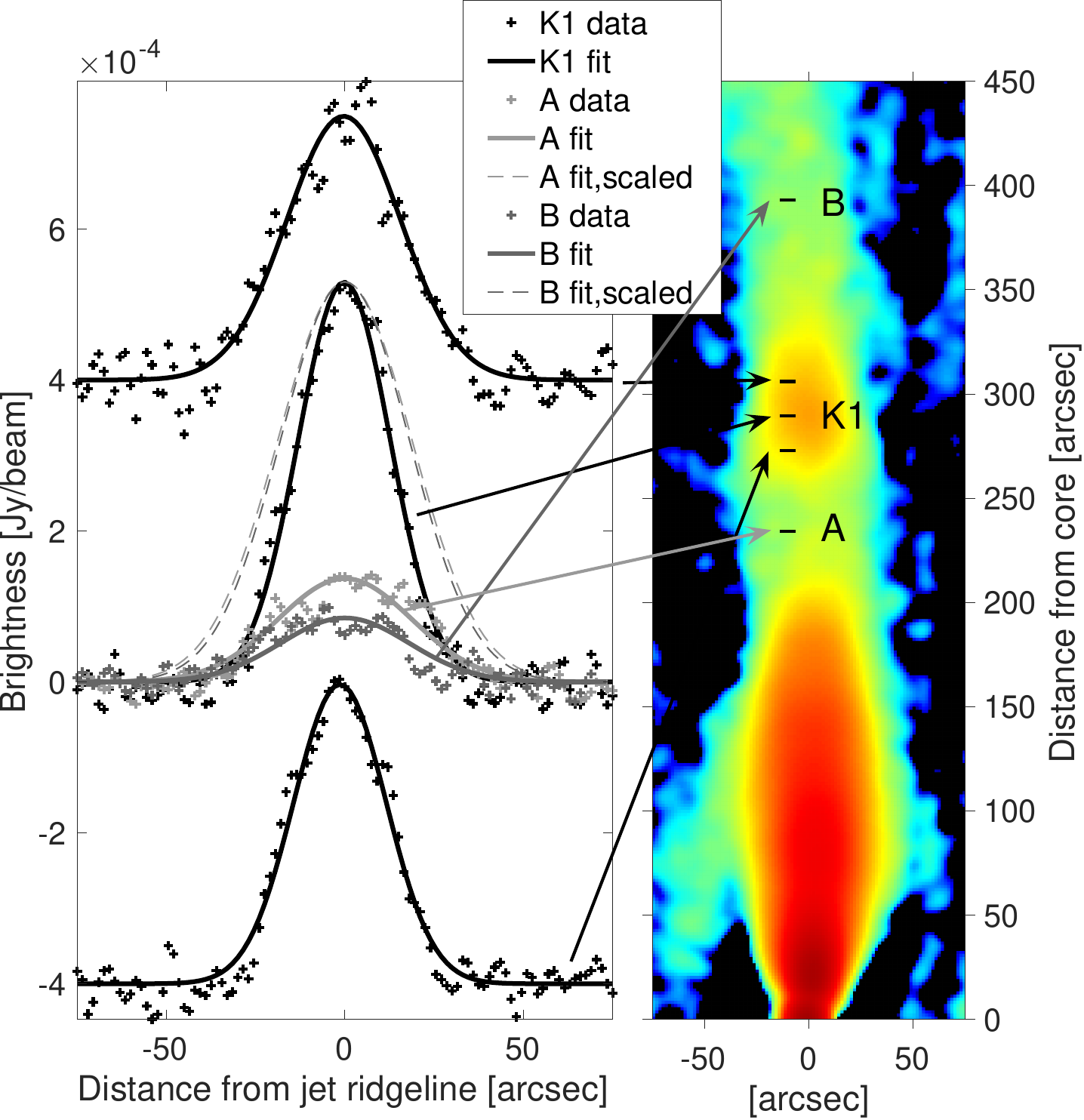}
    \caption{Cross sections of the Southern jet of \ngc\, at various places in the jet. The black crosses show observations of the bright knot K1, for locations of the slices shown in the right panel. Solid lines show one-Gaussian fits to the data. The northern and southern black crosses and lines are offset from zero by $4 \times 10^{-4}$Jy/beam. The dark grey crosses (data) and line (fit) shows the jet upstream from the knot (labelled A). The light grey crosses (data) and line (fit) shows the jet downstream from the knot (labelled B). The grey dashed lines are the same as the grey solid lines, except that they are vertically scaled to the same height as the black line.}
    \label{plot:jetslices}
\end{figure}

Figure \ref{plot:jetslices} shows slices in brightness across various sections of the jet, each 16 arcsec thick. Comparing K1 (black) with rescaled versions of the jet upstream (dark grey) and downstream (light grey) show that the bright knot is \emph{not} just a brighter version of the background jet --- it is narrower. The black lines do not show an obvious spine-sheath structure, as a single Gaussian is a reasonable fit. Attempts at a two Gaussian fit did not find a second, wider underlying Gaussian. At most, the bottom black data shows a narrower peak and broader sides than the single Gaussian, which may indicate a more complicated structure.

Figure \ref{fig_polarisation} (left) shows the fractional polarisation image of \ngc\, which shows that the linear polarisation \emph{decreases} towards the edge of the jet. Previous studies of polarisation inside the jets have shown that fractional polarisation tends to increase near the jet edge (e.g. \citealp{2013AJ....145...49R}, \citealp{2002MNRAS.331..717L}), i.e. the opposite of what we see here.
This most likely indicates an unusual magnetic field in the jet, possibly with an inner coherent field giving way to a mixed, more complex field in the outer sheath. This may indicate the confinement of the jet by large-scale magnetic fields, rather than purely thermal pressure of the environment.

In summary, our observations of the jet are consistent with a core-sheath structure, but this interpretation is far from demanded by the data. A more complete model of this scenario will need to explain why the spine brightens.

\subsection{Bright knots: varying jet power?}
 \label{sec:vjp}
 
If the power of the jet source varies significantly, then past periods of high output will appear later as brighter spots in the jet. If the jets are approximately symmetric at their source, then a bright spot should appear in both jets. However, projection and light travel effects mean that the spots will appear at different distances from the core.
\begin{figure}
    \centering
        \includegraphics[width=0.7\columnwidth]{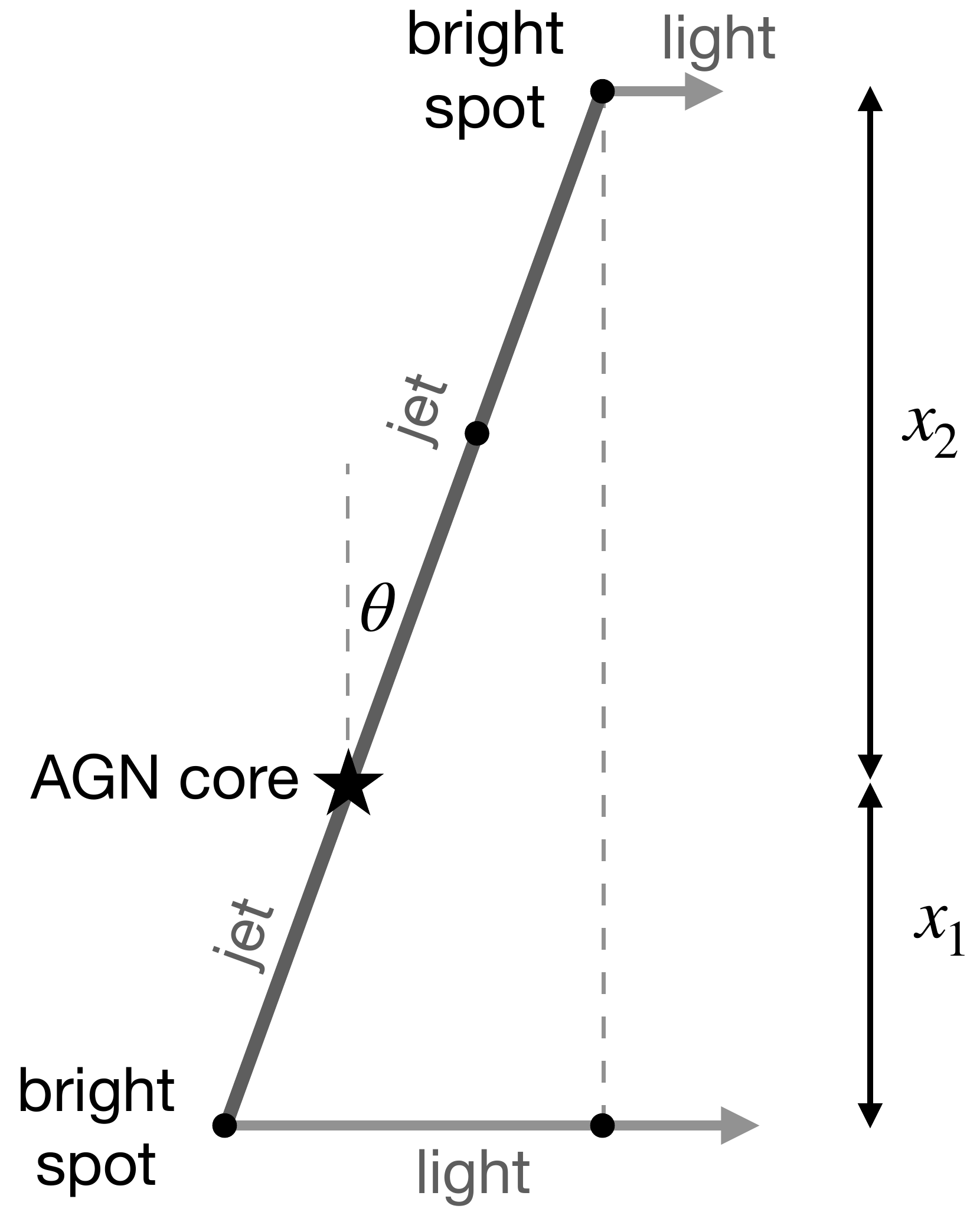}
        \caption{The effect of light travel time on the apparent position of two bright spots. The observer on Earth is located to the right, and the jet is orientated at an angle $\theta$ to the plane of the sky. 
        Jet material is ejected from the \ac{AGN} core at the same time and same speed in opposite directions.
        The relationship between the angle $\theta$, the jet velocity $\beta_{\textrm{jet}}$, and the ratio $x_2 / x_1$ is given in Equation \eqref{eq:jetlighttravel}.}
\label{fig:jetlighttravel}
\end{figure}

Fig.~\ref{fig:jetlighttravel} shows the geometry of the projection effect.

Jet material is ejected from the \ac{AGN} core at the same time and same speed ($\beta_{\textrm{jet}} = v_{\textrm{jet}} / c$) in opposite directions. The lower jet material travels to apparent position $x_1$ and emits light toward the observer. The upper jet material travels somewhat further to $x_2$, before also emitting light toward the observer. These light beams arrive at the observer at the same time. The relationship between the angle $\theta$, the jet velocity $\beta_{\textrm{jet}}$, and the ratio $x_2 / x_1$ is,
\begin{equation} \label{eq:jetlighttravel}
    \frac{x_2}{x_1} = \frac{1 + \beta_{\textrm{jet}} \sin \theta}{1 - \beta_{\textrm{jet}} \sin \theta}
\end{equation}

We can see in Fig.~\ref{fig:pol_ASKAP} that the southern jet (with the recollimation spot) has a larger RM, indicating that its light has passed through a larger column of magnetoionic medium \citep{1988Natur.331..147G}. These electrons are plausibly in the immediate environment of the jet; thus, the southern jet is oriented away from us. It follows that, if there is a corresponding bright spot on the northern jet, it will appear to be \emph{further} away from the core than bright spots in the southern jet. Unfortunately, given the faintness of the northern jet, evidence for bright spots is marginal at best.

This scenario might explain why the northern jet, which seems to be pointing towards us, is fainter than the southern jet. The northern jet is older; or, more precisely, we are observing the northern jet at a later time than the southern jet. If the jet is fading with time, then this will more significantly affect our observations of the northern jet. However, without a specific model of the structure and evolution of the jet, it is difficult to be precise about the magnitude of this effect, relative to relativistic beaming, absorption by the intervening medium, and the asymmetry of the environment.

Further, it is not clear whether periods of higher output from the jet would correspond to a \emph{narrower} bright spot than the rest of the jet. Unless the extra power also narrows the opening angle of the jet, we might expect more energetic regions of the outflow to be less affected by their environment as they travel, and so be broader than the less energetic regions. This is marginal evidence for a recollimation scenario over varying jet power. Exploring this scenario would be aided by simulations, and could be analysed statistically based on a large sample of jets.

\section{Detectability with Distance}
\label{distance}

As noted in the introduction, \ac{AGN} feedback plays a crucial role in galaxy formation, impacting the flow of potentially star-forming gas around large galaxies. The \ac{EMU} survey has and will continue to observe sources such as \ngc, which offer an opportunity to directly observe the effect of \ac{AGN} jets on their environment, and vice versa.

To anticipate the results of the wider EMU survey, we calculate what \ngc\  would look like if it were situated at a range of redshifts. Given that \ngc\ is located at $z_0 = 0.007$, we produce its image at redshift $z$ as follows.
\begin{itemize}
    \item Parts of the original image that are below the one-sigma noise level are removed to produce an image that approximates the ``signal''. 
    \item Pixelate the image to increase the physical resolution scale by $r_A(z) / r_A(z_0)$, where $r_A$ is the angular diameter distance. 
    \item Rescale the brightness of each pixel proportional to \mbox{$(1+z)^{-(3+\alpha)}$}, where the spectral index is taken to be constant $\alpha = -0.7$.
    \item Add Gaussian noise with the same amplitude as the original image, with a spatial scale given by the new pixel size.
\end{itemize}

\begin{figure}
    \centering
    \includegraphics[width=\columnwidth]{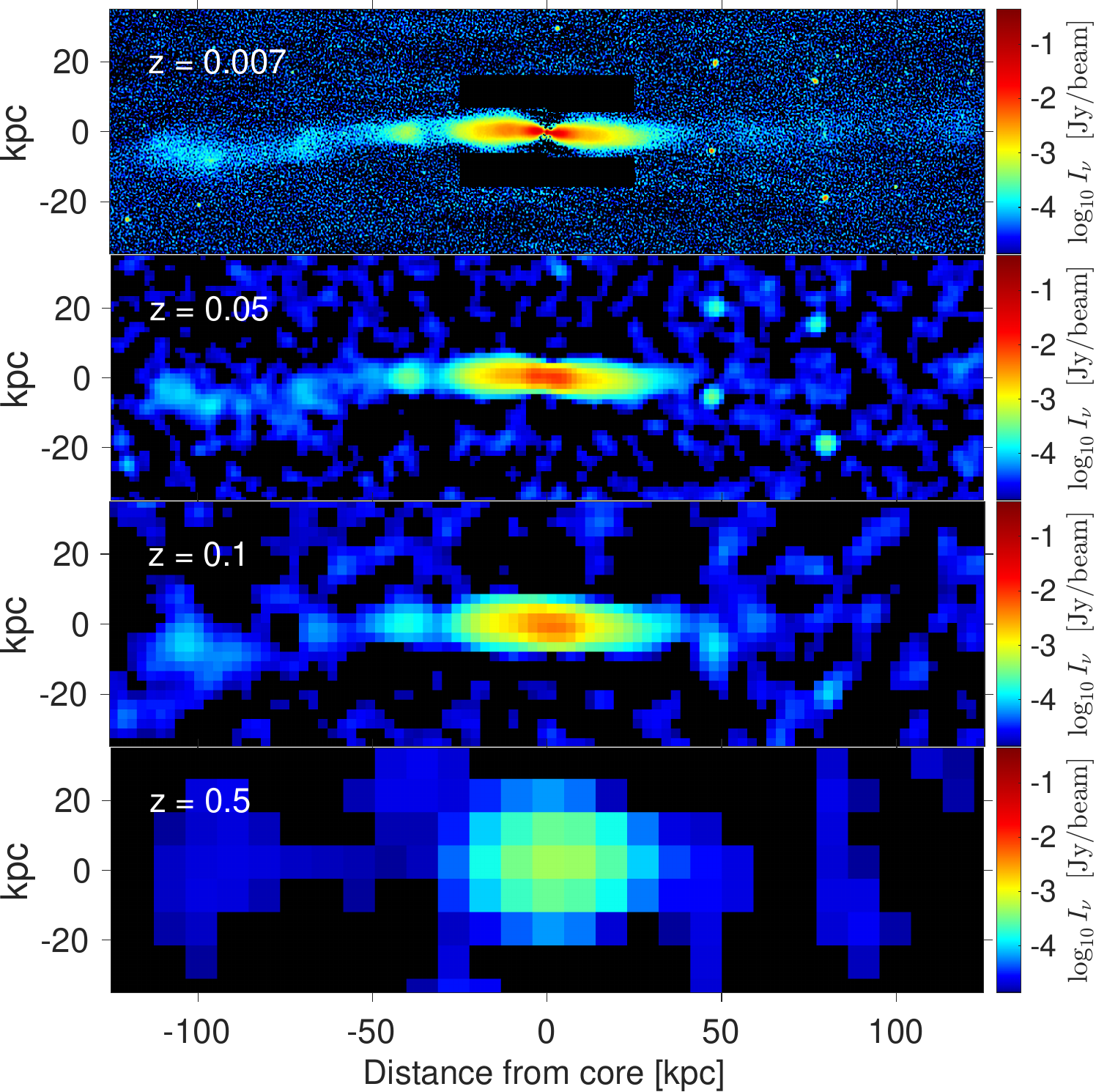}
    \caption{Simulated \ac{ASKAP} images of \ngc, observed as if the source was placed at the redshift shown in each panel. The top panel is the original \ac{ASKAP} brightness image, without any smoothing, and with side lobes above and below the jet are masked.}
    \label{plot:Jet_vs_redshift}
\end{figure}

Fig.~\ref{plot:Jet_vs_redshift} shows simulated images of \ngc, observed as if the source was placed at the redshift shown in each panel. In the top panel (the original image), side lobes above and below the jet are masked. The broad features of the jet are still visible at $z = 0.05$, but by $z = 0.1$ the evidence for a brightening of the southern jet is marginal at best. At $z = 0.5$, the source is detectable but a jet cannot be identified.

The recently completed \ac{EMU} pilot survey, which presents only 1.3~per~cent of the total area that will be covered by the full \ac{EMU} survey \citep{norris2021evolutionary}, has revealed a number of excellent candidates for recollimating \ac{AGN} jets. The \ac{EMU} project is a new-generation radio survey observing the entire southern sky of Dec$<$0 degrees, using \ac{ASKAP}. It covers approximately one-hundredth of area of the planned total \ac{EMU} survey. In addition, the volume of the universe out to $z = 0.1$ is about $10^3$ times larger than the volume out to \ngc. We expect future observations from \ac{EMU} as well as the ongoing VLA Sky Survey \citep[VLASS;][]{2020PASP..132c5001L} and LOFAR Two-Metre Sky Survey \citep[LoTSS][]{2022A&A...659A...1S} will provide many more examples of resolved \ac{AGN} jets \citep[e.g.][]{2019MNRAS.488.2701M}, with which we can test our understanding of their physics and effect on their environment.

\section{Conclusions}

In this paper we report the discovery of one of the largest \ac{AGN} jets in the nearby Universe, associated with the elliptical galaxy \ngc. In summary, we found that:
\begin{itemize}
  \item Radio observations with \ac{ASKAP}, \ac{MWA} and \ac{ATCA} reveal two oppositely directed radio jets that span 355$~$kpc around the otherwise typical red-and-dead elliptical galaxy \ngc. 
  \item X-ray observations with eROSITA, \cxo\ and \Swift\ show that \ngc\ is located in a rarefied environment, with no diffuse emission. 
  \item The jet is remarkably linear over its entire length, which is also consistent with a low density environment on Mpc scales.
  \item The southern jet shows at least one possible example of narrowing and simultaneous brightening of the jet. This is possibly indicative of recollimation of the jet by external, environmental pressure. The large recollimation scale (40$~$kpc) would be consistent with a slow jet in a low pressure environment.
  \item Our observations of the jet could also be explained by a jet with variable power; the brightness of the jet is a record of the past activity of the source. However, it is not clear that this would also produce a narrower region of the jet.
  \item The fractional polarisation of radio emission shows a behaviour uncommon for \ac{AGN} jets: instead of increasing at the edges, it decreases. This may suggest a complicated internal jet structure, rather than external recollimation. 
\end{itemize}

\section*{Acknowledgements}

The \ac{ATCA} and the Australian SKA Pathfinder (ASKAP) are part of the Australian Telescope which is funded by the Commonwealth of Australia for operation as National Facility managed by CSIRO. This paper includes archived data obtained through the Australia Telescope Online Archive (\url{http://atoa.atnf.csiro.au}). Operation of \ac{ASKAP} is funded by the Australian Government with support from the National Collaborative Research Infrastructure Strategy (NCRIS). \ac{ASKAP} uses the resources of the Pawsey Supercomputing Centre. Establishment of \ac{ASKAP}, the Murchison Radio-astronomy Observatory and the Pawsey Supercomputing Centre are initiatives of the Australian Government, with support from the Government of Western Australia and the Science and Industry Endowment Fund.  This scientific work makes use of the Murchison Radio-astronomy Observatory, operated by CSIRO. We acknowledge the Wajarri Yamatji people as the traditional owners of the Observatory site. Support for the operation of the \ac{MWA} is provided by the Australian Government (NCRIS), under a contract to Curtin University administered by Astronomy Australia Limited. The National Radio Astronomy Observatory is a facility of the National Science Foundation operated under cooperative agreement by Associated Universities, Inc. This work made use of the Swinburne University of Technology software correlator, developed as part of the Australian Major National Research Facilities Programme and operated under licence. The recent VLBA experiments were run by the geodetic group of the US Naval Observatory (USNO) to monitor the radio reference frame sources with 10 VLBA antennas \cite{2021AJ....162..121H}.

This work is based on data from eROSITA, the soft X-ray instrument aboard SRG, a joint Russian-German science mission supported by the Russian Space Agency (Roskosmos), in the interests of the Russian Academy of Sciences represented by its Space Research Institute (IKI), and the Deutsches Zentrum für Luft- und Raumfahrt (DLR). The SRG spacecraft was built by Lavochkin Association (NPOL) and its subcontractors, and is operated by NPOL with support from the Max Planck Institute for Extraterrestrial Physics (MPE). The development and construction of the eROSITA X-ray instrument was led by MPE, with contributions from the Dr. Karl Remeis Observatory Bamberg \& ECAP (FAU Erlangen-N\"urnberg), the University of Hamburg Observatory, the Leibniz Institute for Astrophysics Potsdam (AIP), and the Institute for Astronomy and Astrophysics of the University of Tübingen, with the support of DLR and the Max Planck Society. The Argelander Institute for Astronomy of the University of Bonn and the Ludwig Maximilians Universität Munich also participated in the science preparation for eROSITA. The eROSITA data shown here were processed using the eSASS software system developed by the German eROSITA consortium.

H.A. has benefited from grant CIIC 174/2021 of Universidad de Guanajuato, Mexico. 
Partial support for L.R. comes from U.S. National Science Foundation grant AST17-14205 to the University of Minnesota.

We thank Natasha Hurley-Walker for providing the 200-MHz data from the Murchison Widefield Array and Robert Laing for useful discussions and insights which greatly contributed to this paper.

We are also grateful to the anonymous referee for a constructive report and useful comments that helped us improve the paper significantly.

\section*{Data Availability}

The data underlying this article will be shared on reasonable request to the corresponding author.



\bibliographystyle{mnras}
\bibliography{ref} 

\begin{thebibliography}{}
\makeatletter
\relax
\def\mn@urlcharsother{\let\do\@makeother \do\$\do\&\do\#\do\^\do\_\do\%\do\~}
\def\mn@doi{\begingroup\mn@urlcharsother \@ifnextchar [ {\mn@doi@}
  {\mn@doi@[]}}
\def\mn@doi@[#1]#2{\def\@tempa{#1}\ifx\@tempa\@empty \href
  {http://dx.doi.org/#2} {doi:#2}\else \href {http://dx.doi.org/#2} {#1}\fi
  \endgroup}
\def\mn@eprint#1#2{\mn@eprint@#1:#2::\@nil}
\def\mn@eprint@arXiv#1{\href {http://arxiv.org/abs/#1} {{\tt arXiv:#1}}}
\def\mn@eprint@dblp#1{\href {http://dblp.uni-trier.de/rec/bibtex/#1.xml}
  {dblp:#1}}
\def\mn@eprint@#1:#2:#3:#4\@nil{\def\@tempa {#1}\def\@tempb {#2}\def\@tempc
  {#3}\ifx \@tempc \@empty \let \@tempc \@tempb \let \@tempb \@tempa \fi \ifx
  \@tempb \@empty \def\@tempb {arXiv}\fi \@ifundefined
  {mn@eprint@\@tempb}{\@tempb:\@tempc}{\expandafter \expandafter \csname
  mn@eprint@\@tempb\endcsname \expandafter{\@tempc}}}

\bibitem[\protect\citeauthoryear{{Alexander}}{{Alexander}}{2006}]{2006MNRAS.368.1404A}
{Alexander} P.,  2006, \mn@doi [\mnras] {10.1111/j.1365-2966.2006.10225.x},
  \href {https://ui.adsabs.harvard.edu/abs/2006MNRAS.368.1404A} {368, 1404}

\bibitem[\protect\citeauthoryear{{Anderson}, {Heald}, {Eilek}, {Lenc},
  {Gaensler}  et~al.}{{Anderson} et~al.}{2021}]{2021arXiv210201702A}
{Anderson} C.~S.,  {Heald} G.~H.,  {Eilek} J.~A.,  {Lenc} E.,  {Gaensler}
  B.~M.,   et~al., 2021, arXiv e-prints, \href
  {https://ui.adsabs.harvard.edu/abs/2021arXiv210201702A} {p. arXiv:2102.01702}

\bibitem[\protect\citeauthoryear{{Arnaud}}{{Arnaud}}{1996}]{1996ASPC..101...17A}
{Arnaud} K.~A.,  1996, in {Jacoby} G.~H.,  {Barnes} J.,  eds,  Astronomical
  Society of the Pacific Conference Series Vol. 101, Astronomical Data Analysis
  Software and Systems V. p.~17

\bibitem[\protect\citeauthoryear{{Asada}, {Nakamura}, {Doi}, {Nagai}  \&
  {Inoue}}{{Asada} et~al.}{2014}]{2014ApJ...781L...2A}
{Asada} K.,  {Nakamura} M.,  {Doi} A.,  {Nagai} H.,   {Inoue} M.,  2014,
  \mn@doi [\apjl] {10.1088/2041-8205/781/1/L2}, \href
  {https://ui.adsabs.harvard.edu/abs/2014ApJ...781L...2A} {781, L2}

\bibitem[\protect\citeauthoryear{{Barnes} et~al.,}{{Barnes}
  et~al.}{2001}]{2001MNRAS.322..486B}
{Barnes} D.~G.,  et~al., 2001, \mn@doi [\mnras]
  {10.1046/j.1365-8711.2001.04102.x}, \href
  {https://ui.adsabs.harvard.edu/abs/2001MNRAS.322..486B} {322, 486}

\bibitem[\protect\citeauthoryear{{Bertin}, {Mellier}, {Radovich}, {Missonnier},
  {Didelon}  \& {Morin}}{{Bertin} et~al.}{2002}]{2002ASPC..281..228B}
{Bertin} E.,  {Mellier} Y.,  {Radovich} M.,  {Missonnier} G.,  {Didelon} P.,
  {Morin} B.,  2002, in {Bohlender} D.~A.,  {Durand} D.,   {Handley} T.~H.,
  eds,  Astronomical Society of the Pacific Conference Series Vol. 281,
  Astronomical Data Analysis Software and Systems XI. p.~228

\bibitem[\protect\citeauthoryear{{Bicknell}}{{Bicknell}}{1994}]{1994ApJ...422..542B}
{Bicknell} G.~V.,  1994, \mn@doi [\apj] {10.1086/173748}, \href
  {https://ui.adsabs.harvard.edu/abs/1994ApJ...422..542B} {422, 542}

\bibitem[\protect\citeauthoryear{{Boccardi} et~al.,}{{Boccardi}
  et~al.}{2021}]{2021A&A...647A..67B}
{Boccardi} B.,  et~al., 2021, \mn@doi [\aap] {10.1051/0004-6361/202039612},
  \href {https://ui.adsabs.harvard.edu/abs/2021A&A...647A..67B} {647, A67}

\bibitem[\protect\citeauthoryear{{Breeveld} et~al.,}{{Breeveld}
  et~al.}{2010}]{2010MNRAS.406.1687B}
{Breeveld} A.~A.,  et~al., 2010, \mn@doi [\mnras]
  {10.1111/j.1365-2966.2010.16832.x}, \href
  {https://ui.adsabs.harvard.edu/abs/2010MNRAS.406.1687B} {406, 1687}

\bibitem[\protect\citeauthoryear{{Brentjens} \& {de Bruyn}}{{Brentjens} \& {de
  Bruyn}}{2005}]{2005A&A...441.1217B}
{Brentjens} M.~A.,  {de Bruyn} A.~G.,  2005, \mn@doi [\aap]
  {10.1051/0004-6361:20052990}, \href
  {https://ui.adsabs.harvard.edu/abs/2005A&A...441.1217B} {441, 1217}

\bibitem[\protect\citeauthoryear{{Briggs}}{{Briggs}}{1995}]{1995AAS...18711202B}
{Briggs} D.~S.,  1995, in American Astronomical Society Meeting Abstracts. p.
  112.02

\bibitem[\protect\citeauthoryear{{Brunner} et~al.,}{{Brunner}
  et~al.}{2021}]{2021arXiv210614517B}
{Brunner} H.,  et~al., 2021, arXiv e-prints, \href
  {https://ui.adsabs.harvard.edu/abs/2021arXiv210614517B} {p. arXiv:2106.14517}

\bibitem[\protect\citeauthoryear{{Burrows} et~al.,}{{Burrows}
  et~al.}{2005}]{2005SSRv..120..165B}
{Burrows} D.~N.,  et~al., 2005, \mn@doi [\ssr] {10.1007/s11214-005-5097-2},
  \href {https://ui.adsabs.harvard.edu/abs/2005SSRv..120..165B} {120, 165}

\bibitem[\protect\citeauthoryear{{Cantwell} et~al.,}{{Cantwell}
  et~al.}{2020}]{2020MNRAS.495..143C}
{Cantwell} T.~M.,  et~al., 2020, \mn@doi [\mnras] {10.1093/mnras/staa1160},
  \href {https://ui.adsabs.harvard.edu/abs/2020MNRAS.495..143C} {495, 143}

\bibitem[\protect\citeauthoryear{{Cardelli}, {Clayton}  \& {Mathis}}{{Cardelli}
  et~al.}{1989}]{1989ApJ...345..245C}
{Cardelli} J.~A.,  {Clayton} G.~C.,   {Mathis} J.~S.,  1989, \mn@doi [\apj]
  {10.1086/167900}, \href
  {https://ui.adsabs.harvard.edu/abs/1989ApJ...345..245C} {345, 245}

\bibitem[\protect\citeauthoryear{{Cash}}{{Cash}}{1979}]{1979ApJ...228..939C}
{Cash} W.,  1979, \mn@doi [\apj] {10.1086/156922}, \href
  {https://ui.adsabs.harvard.edu/abs/1979ApJ...228..939C} {228, 939}

\bibitem[\protect\citeauthoryear{{Charlot} et~al.,}{{Charlot}
  et~al.}{2020}]{2020A&A...644A.159C}
{Charlot} P.,  et~al., 2020, \mn@doi [\aap] {10.1051/0004-6361/202038368},
  \href {https://ui.adsabs.harvard.edu/abs/2020A&A...644A.159C} {644, A159}

\bibitem[\protect\citeauthoryear{{Cluver}, {Jarrett}, {Dale}, {Smith}, {August}
   \& {Brown}}{{Cluver} et~al.}{2017}]{2017ApJ...850...68C}
{Cluver} M.~E.,  {Jarrett} T.~H.,  {Dale} D.~A.,  {Smith} J. D.~T.,  {August}
  T.,   {Brown} M.~J.~I.,  2017, \mn@doi [\apj] {10.3847/1538-4357/aa92c7},
  \href {https://ui.adsabs.harvard.edu/abs/2017ApJ...850...68C} {850, 68}

\bibitem[\protect\citeauthoryear{{Condon}, {Cotton}, {White}, {Legodi},
  {Goedhart}, {McAlpine}, {Ratcliffe}  \& {Camilo}}{{Condon}
  et~al.}{2021}]{2021ApJ...917...18C}
{Condon} J.~J.,  {Cotton} W.~D.,  {White} S.~V.,  {Legodi} S.,  {Goedhart} S.,
  {McAlpine} K.,  {Ratcliffe} S.~M.,   {Camilo} F.,  2021, \mn@doi [\apj]
  {10.3847/1538-4357/ac0880}, \href
  {https://ui.adsabs.harvard.edu/abs/2021ApJ...917...18C} {917, 18}

\bibitem[\protect\citeauthoryear{{Croton} et~al.,}{{Croton}
  et~al.}{2006}]{2006MNRAS.365...11C}
{Croton} D.~J.,  et~al., 2006, \mn@doi [\mnras]
  {10.1111/j.1365-2966.2005.09675.x}, \href
  {https://ui.adsabs.harvard.edu/abs/2006MNRAS.365...11C} {365, 11}

\bibitem[\protect\citeauthoryear{{Daly} \& {Marscher}}{{Daly} \&
  {Marscher}}{1988}]{1988ApJ...334..539D}
{Daly} R.~A.,  {Marscher} A.~P.,  1988, \mn@doi [\apj] {10.1086/166858}, \href
  {https://ui.adsabs.harvard.edu/abs/1988ApJ...334..539D} {334, 539}

\bibitem[\protect\citeauthoryear{{Danziger} \& {Goss}}{{Danziger} \&
  {Goss}}{1983}]{1983MNRAS.202..703D}
{Danziger} I.~J.,  {Goss} W.~M.,  1983, \mn@doi [\mnras]
  {10.1093/mnras/202.3.703}, \href
  {https://ui.adsabs.harvard.edu/abs/1983MNRAS.202..703D} {202, 703}

\bibitem[\protect\citeauthoryear{{Franceschini} et~al.,}{{Franceschini}
  et~al.}{2003}]{2003MNRAS.343.1181F}
{Franceschini} A.,  et~al., 2003, \mn@doi [\mnras]
  {10.1046/j.1365-8711.2003.06744.x}, \href
  {https://ui.adsabs.harvard.edu/abs/2003MNRAS.343.1181F} {343, 1181}

\bibitem[\protect\citeauthoryear{{Fruscione} et~al.,}{{Fruscione}
  et~al.}{2006}]{2006SPIE.6270E..1VF}
{Fruscione} A.,  et~al., 2006, in Society of Photo-Optical Instrumentation
  Engineers (SPIE) Conference Series. p. 62701V, \mn@doi{10.1117/12.671760}

\bibitem[\protect\citeauthoryear{{Garrington}, {Leahy}, {Conway}  \&
  {Laing}}{{Garrington} et~al.}{1988}]{1988Natur.331..147G}
{Garrington} S.~T.,  {Leahy} J.~P.,  {Conway} R.~G.,   {Laing} R.~A.,  1988,
  \mn@doi [\nat] {10.1038/331147a0}, \href
  {https://ui.adsabs.harvard.edu/abs/1988Natur.331..147G} {331, 147}

\bibitem[\protect\citeauthoryear{{Gehrels} et~al.,}{{Gehrels}
  et~al.}{2004}]{2004ApJ...611.1005G}
{Gehrels} N.,  et~al., 2004, \mn@doi [\apj] {10.1086/422091}, \href
  {https://ui.adsabs.harvard.edu/abs/2004ApJ...611.1005G} {611, 1005}

\bibitem[\protect\citeauthoryear{{G{\'o}mez}, {Mart{\'{\i}}}, {Marscher},
  {Ib{\'a}{\~n}ez}  \& {Alberdi}}{{G{\'o}mez}
  et~al.}{1997}]{1997ApJ...482L..33G}
{G{\'o}mez} J.~L.,  {Mart{\'{\i}}} J.~M.,  {Marscher} A.~P.,  {Ib{\'a}{\~n}ez}
  J.~M.,   {Alberdi} A.,  1997, \mn@doi [\apjl] {10.1086/310671}, \href
  {https://ui.adsabs.harvard.edu/abs/1997ApJ...482L..33G} {482, L33}

\bibitem[\protect\citeauthoryear{{Gooch}}{{Gooch}}{1995}]{1995ASPC...77..144G}
{Gooch} R.,  1995, in {Shaw} R.~A.,  {Payne} H.~E.,   {Hayes} J.~J.~E.,  eds,
  Astronomical Society of the Pacific Conference Series Vol. 77, Astronomical
  Data Analysis Software and Systems IV. p.~144

\bibitem[\protect\citeauthoryear{{Gregory} \& {Loredo}}{{Gregory} \&
  {Loredo}}{1992}]{1992ApJ...398..146G}
{Gregory} P.~C.,  {Loredo} T.~J.,  1992, \mn@doi [\apj] {10.1086/171844}, \href
  {https://ui.adsabs.harvard.edu/abs/1992ApJ...398..146G} {398, 146}

\bibitem[\protect\citeauthoryear{{Guzman} et~al.,}{{Guzman}
  et~al.}{2019}]{Guzman_Askapsoft}
{Guzman} J.,  et~al., 2019, {ASKAPsoft: ASKAP science data processor software}
  (\mn@eprint {ascl} {1912.003})

\bibitem[\protect\citeauthoryear{{Hada} et~al.,}{{Hada}
  et~al.}{2018}]{2018ApJ...860..141H}
{Hada} K.,  et~al., 2018, \mn@doi [\apj] {10.3847/1538-4357/aac49f}, \href
  {https://ui.adsabs.harvard.edu/abs/2018ApJ...860..141H} {860, 141}

\bibitem[\protect\citeauthoryear{{Hardcastle}}{{Hardcastle}}{2018}]{2018NatAs...2..273H}
{Hardcastle} M.,  2018, \mn@doi [Nature Astronomy] {10.1038/s41550-018-0424-1},
  \href {https://ui.adsabs.harvard.edu/abs/2018NatAs...2..273H} {2, 273}

\bibitem[\protect\citeauthoryear{{Hill} et~al.,}{{Hill}
  et~al.}{2004}]{2004SPIE.5165..217H}
{Hill} J.~E.,  et~al., 2004, in {Flanagan} K.~A.,  {Siegmund} O. H.~W.,  eds,
  Society of Photo-Optical Instrumentation Engineers (SPIE) Conference Series
  Vol. 5165, X-Ray and Gamma-Ray Instrumentation for Astronomy XIII. pp
  217--231, \mn@doi{10.1117/12.505728}

\bibitem[\protect\citeauthoryear{{Hine} \& {Longair}}{{Hine} \&
  {Longair}}{1979}]{1979MNRAS.188..111H}
{Hine} R.~G.,  {Longair} M.~S.,  1979, \mn@doi [\mnras]
  {10.1093/mnras/188.1.111}, \href
  {https://ui.adsabs.harvard.edu/abs/1979MNRAS.188..111H} {188, 111}

\bibitem[\protect\citeauthoryear{{Hotan} et~al.,}{{Hotan}
  et~al.}{2021}]{ASKAP_System}
{Hotan} A.~W.,  et~al., 2021, \mn@doi [\pasa] {10.1017/pasa.2021.1}, \href
  {https://ui.adsabs.harvard.edu/abs/2021PASA...38....9H} {38, e009}

\bibitem[\protect\citeauthoryear{{Hunt}, {Johnson}, {Cigan}, {Gordon}  \&
  {Spitzak}}{{Hunt} et~al.}{2021}]{2021AJ....162..121H}
{Hunt} L.~R.,  {Johnson} M.~C.,  {Cigan} P.~J.,  {Gordon} D.,   {Spitzak} J.,
  2021, \mn@doi [\aj] {10.3847/1538-3881/ac135d}, \href
  {https://ui.adsabs.harvard.edu/abs/2021AJ....162..121H} {162, 121}

\bibitem[\protect\citeauthoryear{{Hurley-Walker} et~al.,}{{Hurley-Walker}
  et~al.}{2017}]{2017MNRAS.464.1146H}
{Hurley-Walker} N.,  et~al., 2017, \mn@doi [\mnras] {10.1093/mnras/stw2337},
  \href {https://ui.adsabs.harvard.edu/abs/2017MNRAS.464.1146H} {464, 1146}

\bibitem[\protect\citeauthoryear{{Hutschenreuter} et~al.,}{{Hutschenreuter}
  et~al.}{2022}]{2022A&A...657A..43H}
{Hutschenreuter} S.,  et~al., 2022, \mn@doi [\aap]
  {10.1051/0004-6361/202140486}, \href
  {https://ui.adsabs.harvard.edu/abs/2022A&A...657A..43H} {657, A43}

\bibitem[\protect\citeauthoryear{{Jarrett} et~al.,}{{Jarrett}
  et~al.}{2013}]{2013AJ....145....6J}
{Jarrett} T.~H.,  et~al., 2013, \mn@doi [\aj] {10.1088/0004-6256/145/1/6},
  \href {https://ui.adsabs.harvard.edu/abs/2013AJ....145....6J} {145, 6}

\bibitem[\protect\citeauthoryear{{Jarrett}, {Cluver}, {Brown}, {Dale}, {Tsai}
  \& {Masci}}{{Jarrett} et~al.}{2019}]{2019ApJS..245...25J}
{Jarrett} T.~H.,  {Cluver} M.~E.,  {Brown} M.~J.~I.,  {Dale} D.~A.,  {Tsai}
  C.~W.,   {Masci} F.,  2019, \mn@doi [\apjs] {10.3847/1538-4365/ab521a}, \href
  {https://ui.adsabs.harvard.edu/abs/2019ApJS..245...25J} {245, 25}

\bibitem[\protect\citeauthoryear{{Johnston} et~al.,}{{Johnston}
  et~al.}{2008}]{2008ExA....22..151J}
{Johnston} S.,  et~al., 2008, \mn@doi [Experimental Astronomy]
  {10.1007/s10686-008-9124-7}, \href
  {https://ui.adsabs.harvard.edu/abs/2008ExA....22..151J} {22, 151}

\bibitem[\protect\citeauthoryear{Kaastra}{Kaastra}{1992}]{kaastra1992x}
Kaastra J.,  1992, Internal SRON-Leiden Report

\bibitem[\protect\citeauthoryear{{Kalberla}, {Burton}, {Hartmann}, {Arnal},
  {Bajaja}, {Morras}  \& {P{\"o}ppel}}{{Kalberla}
  et~al.}{2005}]{2005A&A...440..775K}
{Kalberla} P.~M.~W.,  {Burton} W.~B.,  {Hartmann} D.,  {Arnal} E.~M.,  {Bajaja}
  E.,  {Morras} R.,   {P{\"o}ppel} W.~G.~L.,  2005, \mn@doi [\aap]
  {10.1051/0004-6361:20041864}, \href
  {https://ui.adsabs.harvard.edu/abs/2005A&A...440..775K} {440, 775}

\bibitem[\protect\citeauthoryear{{Katz-Stone} \& {Rudnick}}{{Katz-Stone} \&
  {Rudnick}}{1997}]{1997ApJ...488..146K}
{Katz-Stone} D.~M.,  {Rudnick} L.,  1997, \mn@doi [\apj] {10.1086/304661},
  \href {https://ui.adsabs.harvard.edu/abs/1997ApJ...488..146K} {488, 146}

\bibitem[\protect\citeauthoryear{{Komissarov} \& {Falle}}{{Komissarov} \&
  {Falle}}{1997}]{1997MNRAS.288..833K}
{Komissarov} S.~S.,  {Falle} S.~A.~E.~G.,  1997, \mn@doi [\mnras]
  {10.1093/mnras/288.4.833}, \href
  {https://ui.adsabs.harvard.edu/abs/1997MNRAS.288..833K} {288, 833}

\bibitem[\protect\citeauthoryear{{Komissarov} \& {Porth}}{{Komissarov} \&
  {Porth}}{2021}]{2021NewAR..9201610K}
{Komissarov} S.,  {Porth} O.,  2021, \mn@doi [\nar]
  {10.1016/j.newar.2021.101610}, \href
  {https://ui.adsabs.harvard.edu/abs/2021NewAR..9201610K} {92, 101610}

\bibitem[\protect\citeauthoryear{{Kormendy} \& {Richstone}}{{Kormendy} \&
  {Richstone}}{1995}]{1995ARA&A..33..581K}
{Kormendy} J.,  {Richstone} D.,  1995, \mn@doi [\araa]
  {10.1146/annurev.aa.33.090195.003053}, \href
  {https://ui.adsabs.harvard.edu/abs/1995ARA%26A..33..581K} {33, 581}

\bibitem[\protect\citeauthoryear{{Krause}, {Alexander}, {Riley}  \&
  {Hopton}}{{Krause} et~al.}{2012}]{2012MNRAS.427.3196K}
{Krause} M.,  {Alexander} P.,  {Riley} J.,   {Hopton} D.,  2012, \mn@doi
  [\mnras] {10.1111/j.1365-2966.2012.21645.x}, \href
  {https://ui.adsabs.harvard.edu/abs/2012MNRAS.427.3196K} {427, 3196}

\bibitem[\protect\citeauthoryear{{Kronberg}}{{Kronberg}}{1994}]{1994RPPh...57..325K}
{Kronberg} P.~P.,  1994, \mn@doi [Reports on Progress in Physics]
  {10.1088/0034-4885/57/4/001}, \href
  {https://ui.adsabs.harvard.edu/abs/1994RPPh...57..325K} {57, 325}

\bibitem[\protect\citeauthoryear{{Lacy} et~al.,}{{Lacy}
  et~al.}{2020}]{2020PASP..132c5001L}
{Lacy} M.,  et~al., 2020, \mn@doi [\pasp] {10.1088/1538-3873/ab63eb}, \href
  {https://ui.adsabs.harvard.edu/abs/2020PASP..132c5001L} {132, 035001}

\bibitem[\protect\citeauthoryear{{Laing} \& {Bridle}}{{Laing} \&
  {Bridle}}{2014}]{2014MNRAS.437.3405L}
{Laing} R.~A.,  {Bridle} A.~H.,  2014, \mn@doi [\mnras]
  {10.1093/mnras/stt2138}, \href
  {https://ui.adsabs.harvard.edu/abs/2014MNRAS.437.3405L} {437, 3405}

\bibitem[\protect\citeauthoryear{{Laing}, {Bridle}, {Cotton}, {Worrall}  \&
  {Birkinshaw}}{{Laing} et~al.}{2008a}]{2008ASPC..386..110L}
{Laing} R.~A.,  {Bridle} A.~H.,  {Cotton} W.~D.,  {Worrall} D.~M.,
  {Birkinshaw} M.,  2008a, in {Rector} T.~A.,  {De Young} D.~S.,  eds,
  Astronomical Society of the Pacific Conference Series Vol. 386, Extragalactic
  Jets: Theory and Observation from Radio to Gamma Ray. p.~110 (\mn@eprint
  {arXiv} {0801.0154})

\bibitem[\protect\citeauthoryear{{Laing}, {Bridle}, {Parma}, {Feretti},
  {Giovannini}, {Murgia}  \& {Perley}}{{Laing}
  et~al.}{2008b}]{2008MNRAS.386..657L}
{Laing} R.~A.,  {Bridle} A.~H.,  {Parma} P.,  {Feretti} L.,  {Giovannini} G.,
  {Murgia} M.,   {Perley} R.~A.,  2008b, \mn@doi [\mnras]
  {10.1111/j.1365-2966.2008.13091.x}, \href
  {https://ui.adsabs.harvard.edu/abs/2008MNRAS.386..657L} {386, 657}

\bibitem[\protect\citeauthoryear{{Liedahl}, {Osterheld}  \&
  {Goldstein}}{{Liedahl} et~al.}{1995}]{1995ApJ...438L.115L}
{Liedahl} D.~A.,  {Osterheld} A.~L.,   {Goldstein} W.~H.,  1995, \mn@doi
  [\apjl] {10.1086/187729}, \href
  {https://ui.adsabs.harvard.edu/abs/1995ApJ...438L.115L} {438, L115}

\bibitem[\protect\citeauthoryear{{Lloyd} \& {Jones}}{{Lloyd} \&
  {Jones}}{2002}]{2002MNRAS.331..717L}
{Lloyd} B.~D.,  {Jones} P.~A.,  2002, \mn@doi [\mnras]
  {10.1046/j.1365-8711.2002.05239.x}, \href
  {https://ui.adsabs.harvard.edu/abs/2002MNRAS.331..717L} {331, 717}

\bibitem[\protect\citeauthoryear{{Magorrian} et~al.,}{{Magorrian}
  et~al.}{1998}]{1998AJ....115.2285M}
{Magorrian} J.,  et~al., 1998, \mn@doi [\aj] {10.1086/300353}, \href
  {https://ui.adsabs.harvard.edu/abs/1998AJ....115.2285M} {115, 2285}

\bibitem[\protect\citeauthoryear{{Makarov}, {Frouard}, {Berghea}, {Rest},
  {Chambers}, {Kaiser}, {Kudritzki}  \& {Magnier}}{{Makarov}
  et~al.}{2017}]{2017ApJ...835L..30M}
{Makarov} V.~V.,  {Frouard} J.,  {Berghea} C.~T.,  {Rest} A.,  {Chambers}
  K.~C.,  {Kaiser} N.,  {Kudritzki} R.-P.,   {Magnier} E.~A.,  2017, \mn@doi
  [\apjl] {10.3847/2041-8213/835/2/L30}, \href
  {https://ui.adsabs.harvard.edu/abs/2017ApJ...835L..30M} {835, L30}

\bibitem[\protect\citeauthoryear{{McConnell} et~al.,}{{McConnell}
  et~al.}{2016}]{McConnell-ASKAP}
{McConnell} D.,  et~al., 2016, \mn@doi [\pasa] {10.1017/pasa.2016.37}, \href
  {https://ui.adsabs.harvard.edu/abs/2016PASA...33...42M} {33, e042}

\bibitem[\protect\citeauthoryear{{Merloni} et~al.,}{{Merloni}
  et~al.}{2012}]{Merloni+2012}
{Merloni} A.,  et~al., 2012, arXiv e-prints, \href
  {https://ui.adsabs.harvard.edu/abs/2012arXiv1209.3114M} {p. arXiv:1209.3114}

\bibitem[\protect\citeauthoryear{{Mewe}, {Gronenschild}  \& {van den
  Oord}}{{Mewe} et~al.}{1985}]{1985A&AS...62..197M}
{Mewe} R.,  {Gronenschild} E.~H.~B.~M.,   {van den Oord} G.~H.~J.,  1985,
  \aaps, \href {https://ui.adsabs.harvard.edu/abs/1985A&AS...62..197M} {62,
  197}

\bibitem[\protect\citeauthoryear{{Mewe}, {Lemen}  \& {van den Oord}}{{Mewe}
  et~al.}{1986}]{1986A&AS...65..511M}
{Mewe} R.,  {Lemen} J.~R.,   {van den Oord} G.~H.~J.,  1986, \aaps, \href
  {https://ui.adsabs.harvard.edu/abs/1986A&AS...65..511M} {65, 511}

\bibitem[\protect\citeauthoryear{{Mingo} et~al.,}{{Mingo}
  et~al.}{2019}]{2019MNRAS.488.2701M}
{Mingo} B.,  et~al., 2019, \mn@doi [\mnras] {10.1093/mnras/stz1901}, \href
  {https://ui.adsabs.harvard.edu/abs/2019MNRAS.488.2701M} {488, 2701}

\bibitem[\protect\citeauthoryear{{Mizuno}, {G{\'o}mez}, {Nishikawa}, {Meli},
  {Hardee}  \& {Rezzolla}}{{Mizuno} et~al.}{2015}]{2015ApJ...809...38M}
{Mizuno} Y.,  {G{\'o}mez} J.~L.,  {Nishikawa} K.-I.,  {Meli} A.,  {Hardee}
  P.~E.,   {Rezzolla} L.,  2015, \mn@doi [\apj] {10.1088/0004-637X/809/1/38},
  \href {https://ui.adsabs.harvard.edu/abs/2015ApJ...809...38M} {809, 38}

\bibitem[\protect\citeauthoryear{{Napier}, {Bagri}, {Clark}, {Rogers},
  {Romney}, {Thompson}  \& {Walker}}{{Napier} et~al.}{1994}]{Napier1994}
{Napier} P.~J.,  {Bagri} D.~S.,  {Clark} B.~G.,  {Rogers} A. E.~E.,  {Romney}
  J.~D.,  {Thompson} A.~R.,   {Walker} R.~C.,  1994, Proceedings of the IEEE,
  82, 658

\bibitem[\protect\citeauthoryear{{Nelson} et~al.,}{{Nelson}
  et~al.}{2019}]{2019MNRAS.490.3234N}
{Nelson} D.,  et~al., 2019, \mn@doi [\mnras] {10.1093/mnras/stz2306}, \href
  {https://ui.adsabs.harvard.edu/abs/2019MNRAS.490.3234N} {490, 3234}

\bibitem[\protect\citeauthoryear{{Norris} et~al.,}{{Norris}
  et~al.}{2021}]{norris2021evolutionary}
{Norris} R.~P.,  et~al., 2021, \mn@doi [\pasa] {10.1017/pasa.2021.42}, \href
  {https://ui.adsabs.harvard.edu/abs/2021PASA...38...46N} {38, e046}

\bibitem[\protect\citeauthoryear{{Offringa} et~al.,}{{Offringa}
  et~al.}{2014}]{2014MNRAS.444..606O}
{Offringa} A.~R.,  et~al., 2014, \mn@doi [\mnras] {10.1093/mnras/stu1368},
  \href {https://ui.adsabs.harvard.edu/abs/2014MNRAS.444..606O} {444, 606}

\bibitem[\protect\citeauthoryear{{Perucho}, {Lobanov}, {Mart{\'\i}}  \&
  {Hardee}}{{Perucho} et~al.}{2006}]{2006A&A...456..493P}
{Perucho} M.,  {Lobanov} A.~P.,  {Mart{\'\i}} J.~M.,   {Hardee} P.~E.,  2006,
  \mn@doi [\aap] {10.1051/0004-6361:20065310}, \href
  {https://ui.adsabs.harvard.edu/abs/2006A&A...456..493P} {456, 493}

\bibitem[\protect\citeauthoryear{Petit \& Luzum}{Petit \&
  Luzum}{2010}]{iers2010}
Petit G.,  Luzum B.,  eds, 2010, IERS Conventions 2010.
IERS Technical Note No.~36

\bibitem[\protect\citeauthoryear{{Poole} et~al.,}{{Poole}
  et~al.}{2008}]{2008MNRAS.383..627P}
{Poole} T.~S.,  et~al., 2008, \mn@doi [\mnras]
  {10.1111/j.1365-2966.2007.12563.x}, \href
  {https://ui.adsabs.harvard.edu/abs/2008MNRAS.383..627P} {383, 627}

\bibitem[\protect\citeauthoryear{{Predehl} et~al.,}{{Predehl}
  et~al.}{2021}]{Predehl+2020}
{Predehl} P.,  et~al., 2021, \mn@doi [\aap] {10.1051/0004-6361/202039313},
  \href {https://ui.adsabs.harvard.edu/abs/2021A&A...647A...1P} {647, A1}

\bibitem[\protect\citeauthoryear{{Ricci}, {Steiner}  \& {Menezes}}{{Ricci}
  et~al.}{2014a}]{2014MNRAS.440.2419R}
{Ricci} T.~V.,  {Steiner} J.~E.,   {Menezes} R.~B.,  2014a, \mn@doi [\mnras]
  {10.1093/mnras/stu441}, \href
  {https://ui.adsabs.harvard.edu/abs/2014MNRAS.440.2419R} {440, 2419}

\bibitem[\protect\citeauthoryear{{Ricci}, {Steiner}  \& {Menezes}}{{Ricci}
  et~al.}{2014b}]{2014MNRAS.440.2442R}
{Ricci} T.~V.,  {Steiner} J.~E.,   {Menezes} R.~B.,  2014b, \mn@doi [\mnras]
  {10.1093/mnras/stu442}, \href
  {https://ui.adsabs.harvard.edu/abs/2014MNRAS.440.2442R} {440, 2442}

\bibitem[\protect\citeauthoryear{{Roberts}, {Wardle}  \& {Marchenko}}{{Roberts}
  et~al.}{2013}]{2013AJ....145...49R}
{Roberts} D.~H.,  {Wardle} J. F.~C.,   {Marchenko} V.~V.,  2013, \mn@doi [\aj]
  {10.1088/0004-6256/145/2/49}, \href
  {https://ui.adsabs.harvard.edu/abs/2013AJ....145...49R} {145, 49}

\bibitem[\protect\citeauthoryear{{Roming} et~al.,}{{Roming}
  et~al.}{2005}]{2005SSRv..120...95R}
{Roming} P. W.~A.,  et~al., 2005, \mn@doi [\ssr] {10.1007/s11214-005-5095-4},
  \href {https://ui.adsabs.harvard.edu/abs/2005SSRv..120...95R} {120, 95}

\bibitem[\protect\citeauthoryear{{Roming} et~al.,}{{Roming}
  et~al.}{2009}]{2009ApJ...690..163R}
{Roming} P.~W.~A.,  et~al., 2009, \mn@doi [\apj] {10.1088/0004-637X/690/1/163},
  \href {https://ui.adsabs.harvard.edu/abs/2009ApJ...690..163R} {690, 163}

\bibitem[\protect\citeauthoryear{{Sadler}, {Jenkins}  \& {Kotanyi}}{{Sadler}
  et~al.}{1989}]{1989MNRAS.240..591S}
{Sadler} E.~M.,  {Jenkins} C.~R.,   {Kotanyi} C.~G.,  1989, \mn@doi [\mnras]
  {10.1093/mnras/240.3.591}, \href
  {https://ui.adsabs.harvard.edu/abs/1989MNRAS.240..591S} {240, 591}

\bibitem[\protect\citeauthoryear{{Sault} \& {Wieringa}}{{Sault} \&
  {Wieringa}}{1994}]{1994A&AS..108..585S}
{Sault} R.~J.,  {Wieringa} M.~H.,  1994, \aaps, \href
  {http://adsabs.harvard.edu/abs/1994A%26AS..108..585S} {108, 585}

\bibitem[\protect\citeauthoryear{{Sault}, {Teuben}  \& {Wright}}{{Sault}
  et~al.}{1995}]{1995ASPC...77..433S}
{Sault} R.~J.,  {Teuben} P.~J.,   {Wright} M.~C.~H.,  1995, in {Shaw} R.~A.,
  {Payne} H.~E.,   {Hayes} J.~J.~E.,  eds,  Astronomical Society of the Pacific
  Conference Series Vol. 77, Astronomical Data Analysis Software and Systems
  IV. p.~433 (\mn@eprint {} {astro-ph/0612759})

\bibitem[\protect\citeauthoryear{{Schinckel}, {Bunton}, {Cornwell}, {Feain}  \&
  {Hay}}{{Schinckel} et~al.}{2012}]{Schinckel-PAF}
{Schinckel} A.~E.,  {Bunton} J.~D.,  {Cornwell} T.~J.,  {Feain} I.,   {Hay}
  S.~G.,  2012, in \procspie. p. 84442A, \mn@doi{10.1117/12.926959}

\bibitem[\protect\citeauthoryear{Schuh \& Behrend}{Schuh \&
  Behrend}{2012}]{Schuh2012}
Schuh H.,  Behrend D.,  2012, \mn@doi [J Geodyn]
  {http://dx.doi.org/10.1016/j.jog.2012.07.007}, 61, 68

\bibitem[\protect\citeauthoryear{{Shimwell} et~al.,}{{Shimwell}
  et~al.}{2022}]{2022A&A...659A...1S}
{Shimwell} T.~W.,  et~al., 2022, \mn@doi [\aap] {10.1051/0004-6361/202142484},
  \href {https://ui.adsabs.harvard.edu/abs/2022A&A...659A...1S} {659, A1}

\bibitem[\protect\citeauthoryear{{Spingola}, {Dallacasa}, {Belladitta},
  {Caccianiga}, {Giroletti}, {Moretti}  \& {Orienti}}{{Spingola}
  et~al.}{2020}]{2020A&A...643L..12S}
{Spingola} C.,  {Dallacasa} D.,  {Belladitta} S.,  {Caccianiga} A.,
  {Giroletti} M.,  {Moretti} A.,   {Orienti} M.,  2020, \mn@doi [\aap]
  {10.1051/0004-6361/202039458}, \href
  {https://ui.adsabs.harvard.edu/abs/2020A&A...643L..12S} {643, L12}

\bibitem[\protect\citeauthoryear{{Thomas}, {Saglia}, {Bender}, {Erwin}  \&
  {Fabricius}}{{Thomas} et~al.}{2014}]{2014ApJ...782...39T}
{Thomas} J.,  {Saglia} R.~P.,  {Bender} R.,  {Erwin} P.,   {Fabricius} M.,
  2014, \mn@doi [\apj] {10.1088/0004-637X/782/1/39}, \href
  {https://ui.adsabs.harvard.edu/abs/2014ApJ...782...39T} {782, 39}

\bibitem[\protect\citeauthoryear{{Tingay} et~al.,}{{Tingay}
  et~al.}{2013}]{2013PASA...30....7T}
{Tingay} S.~J.,  et~al., 2013, \mn@doi [\pasa] {10.1017/pasa.2012.007}, \href
  {https://ui.adsabs.harvard.edu/abs/2013PASA...30....7T} {30, e007}

\bibitem[\protect\citeauthoryear{{Titov}, {Tesmer}  \& {Boehm}}{{Titov}
  et~al.}{2004}]{Titov2004}
{Titov} O.,  {Tesmer} V.,   {Boehm} J.,  2004, in {Vandenberg} N.~R.,  {Baver}
  K.~D.,  eds, International VLBI Service for Geodesy and Astrometry 2004
  General Meeting Proceedings. p.~267

\bibitem[\protect\citeauthoryear{{Veilleux} \& {Osterbrock}}{{Veilleux} \&
  {Osterbrock}}{1987}]{1987ApJS...63..295V}
{Veilleux} S.,  {Osterbrock} D.~E.,  1987, \mn@doi [\apjs] {10.1086/191166},
  \href {https://ui.adsabs.harvard.edu/abs/1987ApJS...63..295V} {63, 295}

\bibitem[\protect\citeauthoryear{{Wayth} et~al.,}{{Wayth}
  et~al.}{2018}]{Wayth-PhaseII}
{Wayth} R.~B.,  et~al., 2018, \mn@doi [\pasa] {10.1017/pasa.2018.37}, \href
  {https://ui.adsabs.harvard.edu/abs/2018PASA...35...33W} {35, 33}

\bibitem[\protect\citeauthoryear{{Willick}, {Courteau}, {Faber}, {Burstein},
  {Dekel}  \& {Strauss}}{{Willick} et~al.}{1997}]{1997ApJS..109..333W}
{Willick} J.~A.,  {Courteau} S.,  {Faber} S.~M.,  {Burstein} D.,  {Dekel} A.,
  {Strauss} M.~A.,  1997, \mn@doi [\apjs] {10.1086/312983}, \href
  {https://ui.adsabs.harvard.edu/abs/1997ApJS..109..333W} {109, 333}

\bibitem[\protect\citeauthoryear{{Wilms}, {Allen}  \& {McCray}}{{Wilms}
  et~al.}{2000}]{2000ApJ...542..914W}
{Wilms} J.,  {Allen} A.,   {McCray} R.,  2000, \mn@doi [\apj] {10.1086/317016},
  \href {https://ui.adsabs.harvard.edu/abs/2000ApJ...542..914W} {542, 914}

\bibitem[\protect\citeauthoryear{{Wykes}, {Hardcastle}, {Karakas}  \&
  {Vink}}{{Wykes} et~al.}{2015}]{2015MNRAS.447.1001W}
{Wykes} S.,  {Hardcastle} M.~J.,  {Karakas} A.~I.,   {Vink} J.~S.,  2015,
  \mn@doi [\mnras] {10.1093/mnras/stu2440}, \href
  {https://ui.adsabs.harvard.edu/abs/2015MNRAS.447.1001W} {447, 1001}

\makeatother
\end{thebibliography}


\bsp	
\label{lastpage}
\end{document}